\documentclass[%
 reprint,
superscriptaddress,
 amsmath,amssymb,
 aps,
]{revtex4}
\usepackage{graphicx}% Include figure files
\usepackage[percent]{overpic}
\usepackage{dcolumn}% Align table columns on decimal point
\usepackage{bm}% bold math
\usepackage[colorlinks,linkcolor=red,anchorcolor=blue,citecolor=green]{hyperref}
\hypersetup{
    colorlinks=true,
    linkcolor=red,
    filecolor=gray,
    urlcolor=blue,
    citecolor=blue,
}

\usepackage{subfigure}
\usepackage{float}
\usepackage{booktabs}
\usepackage{tabularx}

\usepackage{orcidlink}
\usepackage{pifont} %
\usepackage{setspace}

\begin{document}

%\preprint{APS/123-QED}

\title{Periodic orbits around a magnetically charged black hole in $f(R,T)$ gravity coupled with Euler-Heisenberg electrodynamics}% Force line breaks with \\
\author{Jun-Long Huang}

\email{huangjunlong@stu.scu.edu.cn}
\affiliation{College of Physics, Sichuan University, Chengdu, 610065, China}

\author{Chen-Kai Qiao}
\email{chenkaiqiao@cqut.edu.cn}
\affiliation{College of Physical Science and New Energy, Chongqing University of Technology, Chongqing 400054, China}

\author{Jun Tao}
\email{taojun@scu.edu.cn}
\affiliation{College of Physics, Sichuan University, Chengdu, 610065, China}

\date{\today}

\begin{abstract}
In this work, we systematically investigate the zoom-whirl periodic orbits of a small compact object orbiting a magnetically charged black hole in $f(R, T)$ gravity coupled with Euler-Heisenberg nonlinear electrodynamics. 
We evaluate the impact of the magnetic charge $Q_m$ and the Euler-Heisenberg parameter $a$ on the characteristics of the innermost stable circular orbit (ISCO) and the marginally bound orbit (MBO). 
Particularly, we focus on the precession parameter q and systematically examine how the magnetic charge and other parameters affect the trajectories of periodic orbits. 
Using the numerical kludge method, we generate gravitational waveforms for these periodic orbits. 
The results presented in our work demonstrate that the magnetic charge can significantly modify not only the precession parameter but also the orbital trajectories of the periodic orbits. Moreover, changing the magnetic charge $Q_m$ could cause a significant phase shift in the gravitational waveforms, while other parameters exert a relatively weaker influence on the gravitational waves.
\end{abstract}

\maketitle

%\tableofcontents

\section{Introduction}\label{sec1}

Einstein's general relativity serves as the cornerstone of theoretical physics and cosmology. It predicts many novel physical phenomena for gravitational systems. 
Recently, gravitational waves, black hole shadows, and other observations have directly verified its immense success \cite{LIGOScientific:2016aoc,LIGOScientific:2017ync,Will:2014kxa,Ishak:2018his,LIGOScientific:2020kqk,EventHorizonTelescope:2022wkp,EventHorizonTelescope:2019dse}. 
Although multi-messenger astronomical observations do not seem to show significant deviations from general relativity, there are a number of phenomena (such as dark matter, dark energy, the Hubble constant, and primordial gravitational waves) that remain difficult to explain within the framework of general relativity. 
These astrophysical phenomena exhibit features beyond the standard cosmological model or the pure Einstein field equations \cite{Clifton:2011jh,Ezquiaga:2017ekz,Kobayashi:2011nu}. Firstly, to meet observational constraints from cosmology (such as the accelerated expansion of the universe and the inflation in the early universe), the cosmological constant and the matter constitutions must be fine-tuned with remarkable precision to a very narrow range \cite{Weinberg:1988cp,Guth:1980zm,Peebles:2002gy,Planck:2018vyg,Planck:2018jri}. 
This fine-tuning requirement points to the limitations of general relativity and cosmologcial models in explaining the universe at the beginning stage or at extreme scales \cite{Joyce:2014kja}. 
In addition to these cosmological challenges, on galactic scales, observations of galaxy rotation curves and gravitational lensing of bullet cluster suggest that the Newtonian gravity and general relativity predictions also have discrepancies in fitting these galactic observations \cite{Sofue:2000jx,Rubin:1980zd,Clowe:2006eq}. 
To explain these observational results on galactic scales, one must either introduce a substantial amount of invisible dark matter \cite{Saxton:2010jk,Navarro:1996gj} or fundamentally modify the Einstein field equations to avoid relying on the unknown dark matter \cite{Bekenstein:2004ne,Capozziello:2006uv,Verlinde:2016toy,Moffat:2013sja,Mannheim:1988dj}. Furthermore, the non-renormalizability of general relativity in the quantum context also implies that necessary modifications to Einstein's theory may be inevitable \cite{tHooft:1974toh,Niedermaier:2006ns,Goroff:1985th,Horava:2009uw,Arkani-Hamed:2008owk,Bern:2007hh,Witten:2012bh}. In summary, the pursuit of a consistent quantum gravitational theory together with multi-band observational data requires the modification or extension of the Einstein field equations.

Modified gravity theories and nonlinear electromagnetic theories constitute two important frontier directions beyond the standard general relativity and cosmological model \cite{Verlinde:2010hp,AraujoFilho:2025hnf,Li:2025eln,Heisenberg:1936nmg,Harko:2011kv,Carroll:2003wy,Capozziello:2011et,Born:1933pep,Sorokin:2021tge,Li:2025rog}. 
The $f(R, T)$ gravity theory generalizes the Einstein-Hilbert action through an arbitrary function of the Ricci scalar $R$ and the trace $T$ of the energy-momentum tensor \cite{Harko:2011kv,Nojiri:2017ncd,Nojiri:2010wj}. This theory introduces a direct coupling between matter fields and spacetime geometry. On local astrophysical scales \cite{Moraes:2017zgm}, it alters the spacetime structure of compact objects and predicts distinct observational features \cite{Hazarika:2024cji,Asghari:2024qgp}. 
In addition, when considering the interplay between quantum gravity effects and strong electromagnetic fields, one typically requires descriptions of electrodynamics beyond linear theories. 
The Euler-Heisenberg theory \cite{Heisenberg:1936nmg,Ruffini:2013hia,Yajima:2000kw} provides a low-energy effective Lagrangian accounting for quantum electrodynamical effects. It serves as a promising candidate for nonlinear electrodynamics when coupled with gravity systems. Furthermore, combining $f(R, T)$ gravity with nonlinear electrodynamics, novel electromagnetic black hole solutions have been reported recently \cite{Liang:2025hzr,Rois:2025tfe,AraujoFilho:2025hnf}. In these black holes, the spacetime  deviates from the conventional RN spacetime and is characterized by a nonlinear electromagnetic parameter $a$ and a coupling strength $\beta$ between matter fields and spacetime geometry.
These solutions thus provide frameworks for systematically investigating black hole spacetimes under the joint influence of matter-geometry coupling and nonlinear electromagnetic effects.

The periodic orbits are a class of closed bound orbits characterized by zoom-whirl behavior \cite{Glampedakis:2002ya}. 
These orbits can be classified by three integers $(z, w, v)$ \cite{Levin:2008mq,Levin:2008ci,Misra:2010pu}. 
Here, $z$ is the zoom number, which describes the number of ``leaves'' in this periodic orbit; $w$ indicates the number of additional whirls during one complete radial period (in addition to its basic precession motion from apoapsis to periapsis and back to apoapsis, the particle revolves around the center by $w$ additional loops with azimuthal angle of $2\pi w$); and $v$ represents the ``jump'' between apocenters. This kind of periodic orbit has become an inspiring topic and has motivated a large number of research works in recent years \cite{He:2026txc,Babar:2017gsg,Liu:2018vea,Wei:2019zdf,Tu:2023xab,Li:2024tld,Yang:2024lmj,Shabbir:2025kqh,Chen:2025aqh,Hua:2026kvw,Shabbir:2026qlh,Huang:2026oga,Wang:2025wob,Choudhury:2025qsh,Heidari:2026eim,Gogoi:2026obd,Lu:2025xlp,Heidari:2026zff,Deng:2020hxw,Wang:2022tfo,Lin:2023rmo,Alloqulov:2025ucf,Alloqulov:2025bxh,Alloqulov:2025dqi,Ahmed:2025azu,Ahmed:2025shr,Ahmed:2026cgx}. 
Investigating the periodic orbits of companion stars around black holes serves as an essential probe for exploring the spacetime geometry near the central supermassive black hole. 
Furthermore, it provides an avenue to test the validity of modified gravitational theories in the strong-field regime, while also shedding light on the physics and observations in extreme mass ratio inspiral systems (EMRIs). 
Specifically, EMRIs are formed by a supermassive black hole in the galaxy's center and small compact objects revolving around it.
In these systems, the small compact object moves along intricate orbits around the central black hole. Such EMRIs are among the most exciting potential sources of gravitational waves for future low-frequency, space-based detectors, such as the Laser Interferometer Space Antenna (LISA) \cite{LISA:2017pwj}, Taiji \cite{Hu:2017mde}, TianQin \cite{TianQin:2015yph}, and the DECI-hertz Interferometer Gravitational-wave Observatory (DECIGO) \cite{Kawamura:2011zz}. 
These space-based gravitational wave detectors can provide high-precision observational data, which enables the capture of gravitational wave signals from periodic orbits and the identification of the unique zoom-whirl behavior. 

In particular, exploring the impact of modified gravity theories (such as $f(R, T)$ gravity \cite{Harko:2011kv}) and nonlinear electrodynamics (such as Euler-Heisenberg theory \cite{Yajima:2000kw}) on EMRI gravitational wave signals is of great importance. 
To help us better test these theories through gravitational wave detections, it is necessary to explore the black hole solutions in the presence of nonlinear electrodynamics and modified gravity and study the bound orbits in such systems. 
Analyzing how the different model parameters influence bound orbits and associated gravitational waves (especially the zoom-whirl signature of periodic orbits) would be important for probing these theories. 
Under these circumstances, the main purpose of this paper is to investigate the periodic orbits of small compact objects moving around the magnetically charged black hole in the framework of $f(R,T)$ theory coupled with Euler-Heisenberg electrodynamics and extract the corresponding gravitational waveforms of these orbits. Specifically, we employ the numerical kludge method based on the quadrupole moment approximation \cite{Gair:2005is,Babak:2006uv} to generate the gravitational waveforms.

The structure of this paper is organized as follows: In Section \ref{sec2}, we briefly review the spacetime metric of a magnetically charged black hole in $f(R, T)$ gravity coupled with Euler-Heisenberg electrodynamics. 
We present the equation of motion for a small compact object, analyzing the characteristics of the innermost stable circular orbit (ISCO) and the marginally bound orbit (MBO). 
In Section \ref{sec3}, we investigated the precession parameter $q$ and systematically examined how the magnetic charge and other parameters affect the trajectories of bound periodic orbits. By numerically solving the geodesic equations, we explore the dependence of the orbital precession parameter on energy and angular momentum, and present various complex zoom-whirl trajectories. 
In Section \ref{sec4}, we compute the gravitational waveforms generated from these periodic orbits in EMRIs. We investigate how the topological structure of the orbits and the magnetic charge imprint distinct features on the gravitational waveforms. Finally, Section \ref{sec5} summarizes the main conclusions obtained in this paper. For simplicity, we adopt the natural units $G = c = 1$ throughout this work.

\section{magnetically charged black hole in $f(R, T)$ gravity coupled with Euler-Heisenberg electrodynamics}\label{sec2}

In this section, we first briefly review the spacetime metric for a magnetically charged black hole in $f(R, T)$ gravity coupled with Euler-Heisenberg electrodynamics. This black hole solution is derived by Liang et al. in ref.~\cite{Liang:2025hzr}, which suggests the following metric form
\begin{eqnarray}\label{linele}
	d s^2=-f(r) d t^2+\frac{1}{f(r)} d r^2+r^2\left(d \theta^2+\sin ^2 \theta d \phi^2\right),
\end{eqnarray}
with the metric function $f(r)$ depending on the Euler-Heisenberg parameter $a$, coupling strength $\beta$, and the magnetic charge $Q_m$. Following the derivation in ref.~\cite{Liang:2025hzr}, the metric function is given by
\begin{eqnarray}\label{metric}
	f(r) & = & 1 - \frac{2M}{r} + \frac{Q_m^2}{r^2} - \frac{a Q_m^4}{10 r^6}+\frac{\beta a Q_m^4}{20\pi r^6},
\end{eqnarray}
where $M$ is the Arnowitt-Deser-Misner (ADM) mass. For the parameters $a$ and $\beta$, $a$ characterizes the strength of the nonlinear electromagnetic effects, while $\beta$ characterizes the coupling strength between matter and spacetime geometry. Particularly, in the absence of nonlinear electromagnetic effects (with $a=0$), this spacetime recovers the standard Reissner-Nordstr\"om(RN) spacetime with magnetic charge. In the absence of coupling strength $\beta$, the spacetime reduces to the magnetically charged black hole in the Einstein-Euler-Heisenberg theory \cite{Zeng:2022pvb,Magos:2020ykt,Yu:2023glo,Luo:2026srx} \footnote{In some literature, the Euler-Heisenberg parameter $a$ maybe rescaled from our work by $a\to \frac{a}{2}$. In this work, we follow the notation in Liang's work, where the Lagrangian density for Euler-Heisenberg theory is written as $L=\frac{1}{4\pi}(-F+aF^2+bG^2)$ with $b=\frac{7a}{4}$. Some literature adopted the Lagrangian $L=\frac{1}{4\pi}(-F+\frac{a}{2}F^2+\frac{7a}{8}G^2)$}.
In this work, we choose $0 \leq a/M^2 \leq 10$ with the constraint $|2\beta| \leq 2$. Considering the motion of a massive test particle in a black hole background, the Lagrangian of the particle is
\begin{eqnarray}\label{Lagrangian}
	\mathcal{L}=\frac{m}{2}g_{\mu\nu}\frac{dx^{\mu}}{d\tau}\frac{dx^{\nu}}{d\tau},
	\end{eqnarray}
where $\tau$ is the proper time of the test particle, and $m$ is the mass of the test particle. Without any loss of generality, we can conveniently normalize the mass of test particle (or small compact object) as $m=1$, such that the Lagrangian is expressed as $2\mathcal{L}=-1$, and the generalized momentum of the particle is given by
\begin{eqnarray}\label{generalized momentum}
	p_{\mu}=\frac{\partial\mathcal{L}}{\partial\dot{x}^{\mu}}=g_{\mu\nu}\dot{x}^{\nu},
\end{eqnarray}
where the dot denotes the derivative with respect to the proper time. Substituting Eq.~(\ref{generalized momentum}) into Eqs.~(\ref{linele}) and (\ref{metric}). we can derive the equations of generalized momentum for the particle
\begin{subequations}
\begin{eqnarray}\label{four generalized momentum}
	p_t&=&-\left(1 - \frac{2M}{r} + \frac{Q_m^2}{r^2} - \frac{a Q_m^4}{10 r^6}+\frac{\beta a Q_m^4}{20\pi r^6}\right)\dot{t}=-E,\\
	p_r&=&\frac{\dot{r}}{1 - \frac{2M}{r} + \frac{Q_m^2}{r^2} - \frac{a Q_m^4}{10 r^6}+\frac{\beta a Q_m^4}{20\pi r^6}},\\
	p_\theta&=&r^2\dot{\theta},\\
	p_\phi&=&r^2\sin ^2 \theta\cdot\dot{\phi}=L,
\end{eqnarray}	
\end{subequations}
where $E$ and $L$ represent the energy and the orbital angular momentum of the particle per unit mass, respectively. 

\begin{figure}
    \centering
    \begin{minipage}[b]{0.45\textwidth}
        \centering
        \includegraphics[width=\linewidth]{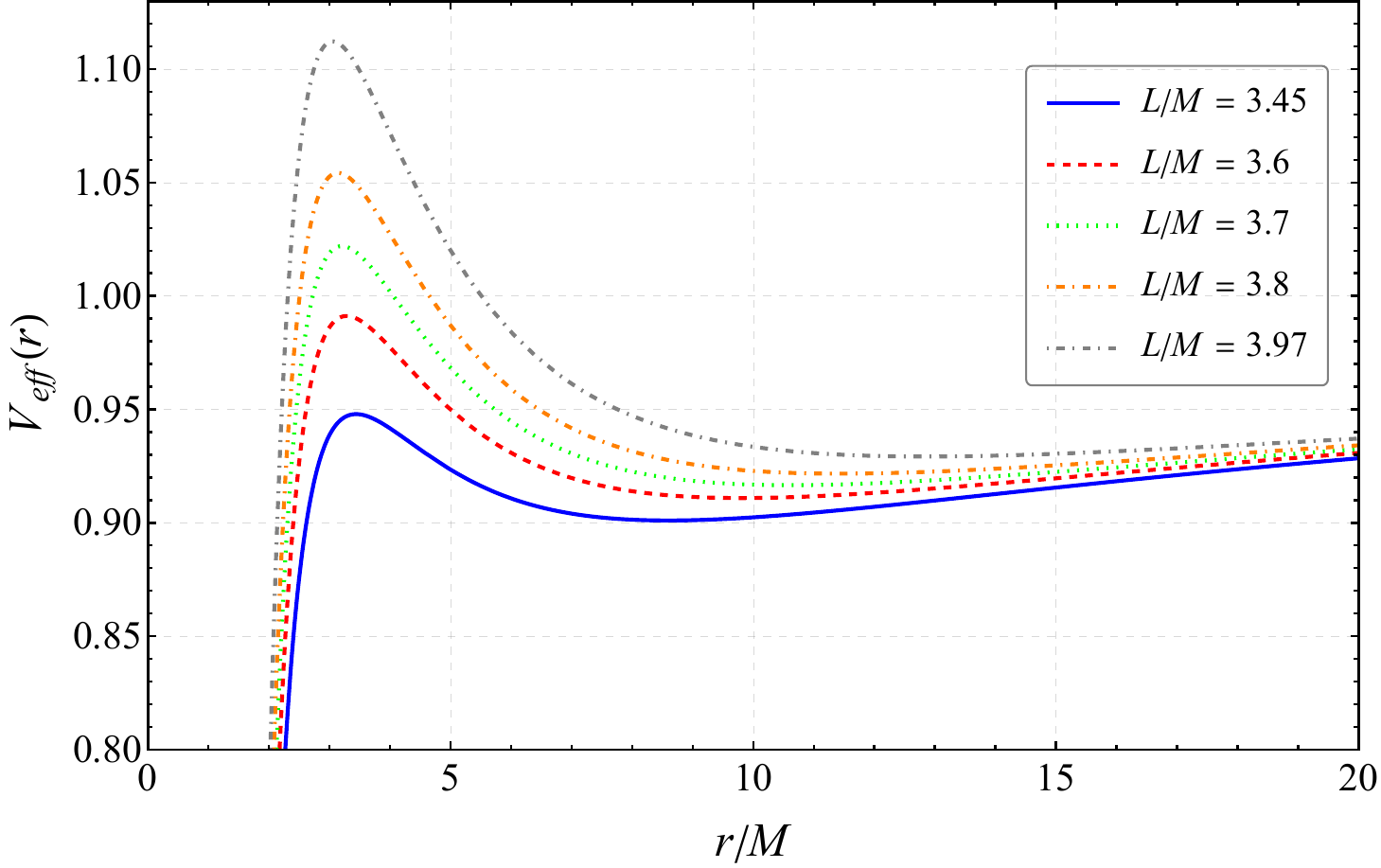}
    \end{minipage}
    %\hfill
    \begin{minipage}[b]{0.45\textwidth}
        \centering
        \includegraphics[width=\linewidth]{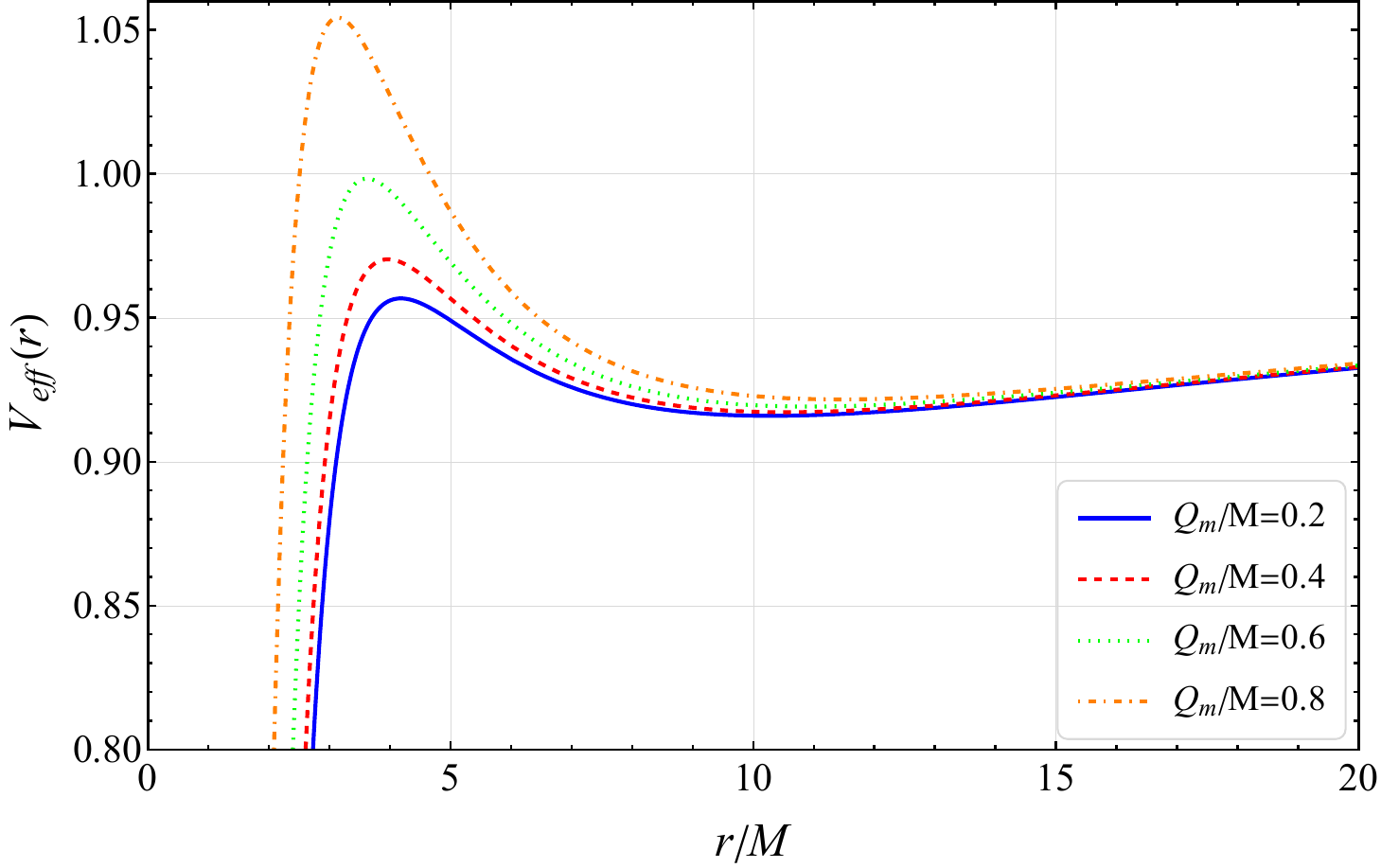}
    \end{minipage}

    \vspace{0.3cm}

    \begin{minipage}[b]{0.45\textwidth}
        \centering
        \includegraphics[width=\linewidth]{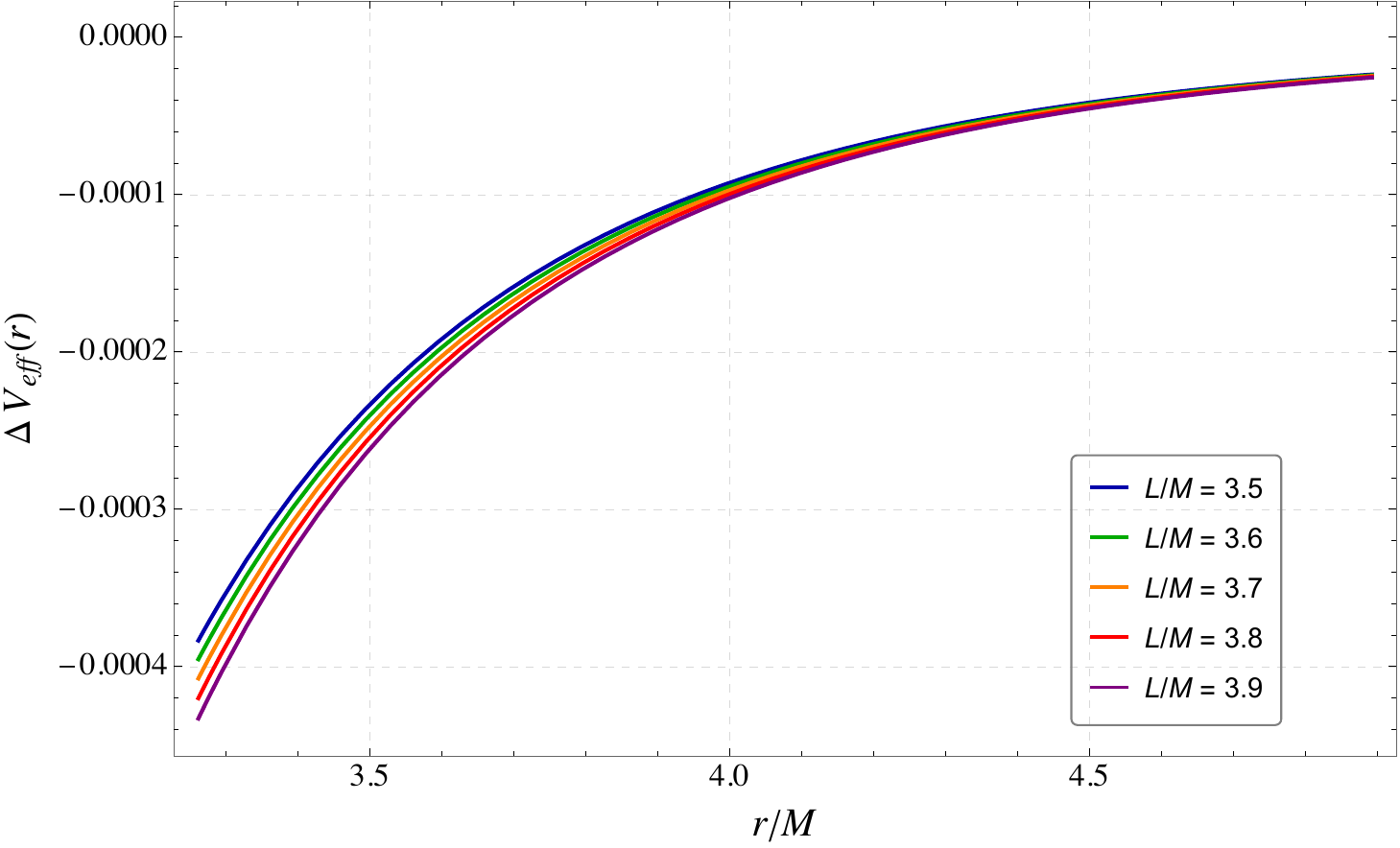}
    \end{minipage}

    \caption{The effective potential $V_{\text{eff}}$ under different magnetic charges and angular momenta. The upper left panel highlights the variational trend of the effective potential for different angular momenta (with the magnetic charge chosen to be $Q_m/M=0.2$). The upper right panel displays the effective potentials for several magnetic charges (with the angular momentum fixed at $L/M = 3.8$). The lower panel shows the difference in the effective potential, $\Delta V_{\rm eff}(r)=V_{\rm eff}(r;a/M^2=8)-V_{\rm eff}(r;a/M^2=2)$, for several fixed angular momenta $L/M=3.5,\,3.6,\,3.7,\,3.8,\,3.9$, with $Q_m/M=0.8$ and $\beta=0.8$. The radial interval is chosen from the minimum value of $r_{\rm ISCO}$ to the maximum value of $r_{\rm MBO}$ obtained for the two cases $a/M^2=2$ and $a/M^2=8$. In this figure, we adopt the parameter values $a/M^2=5$ and $\beta=0.4$ throughout the upper panels.}
    \label{fig:effective_potential}
\end{figure}

Since we consider a spherically symmetric magnetically charged black hole, we can always constrain the motion of test particles in the equatorial plane, so that $\theta = \pi/2$ and $\dot{\theta}=0$, and the Lagrangian in Eq.~(\ref{Lagrangian}) leads to:
\begin{eqnarray}\label{Sphericity metricSphericity metric}
	2\mathcal{L}=-1 \ \Rightarrow \ 
    \frac{\dot{r}^2}{1 - \frac{2M}{r} + \frac{Q_m^2}{r^2} - \frac{a Q_m^4}{10 r^6}+\frac{\beta a Q_m^4}{20\pi r^6}}+\frac{L^2}{r^2}-\frac{E^2}{1 - \frac{2M}{r} + \frac{Q_m^2}{r^2} - \frac{a Q_m^4}{10 r^6}+\frac{\beta a Q_m^4}{20\pi r^6}}=-1.
\end{eqnarray}	
We can rewrite Eq.~(\ref{Sphericity metricSphericity metric}) in the following form
\begin{eqnarray}\label{Simplified metric}
	\dot{r}^2+V_{\text{eff}}=E^2,
\end{eqnarray}
where $V_{\text{eff}}$ is the effective potential
\begin{eqnarray}\label{Effective potential}
	V_{\text{eff}}=\left(1 - \frac{2M}{r} + \frac{Q_m^2}{r^2} - \frac{a Q_m^4}{10 r^6}+\frac{\beta a Q_m^4}{20\pi r^6}\right)\left(1+\frac{L^2}{r^2}\right).
\end{eqnarray}
To intuitively understand the dynamic behavior of the test particle, we plot the effective potential $V_{\text{eff}}$ as a function of the radial coordinate $r$ in Fig.~\ref{fig:effective_potential}. 
It clearly illustrates how the shape of the potential well and the potential barrier are modified by these parameters, which directly determines the allowed range of bound orbits. From the upper panels, it is demonstrated that the potential barrier
 of this magnetically charged black hole in $f(R,T)$ gravity coupled with Euler-Heisenberg electrodynamics can be greatly enhanced when the magnetic charge $Q_m$ takes a larger value. 
 However, the nonlinear electromagnetic effects (parameterized by $a$) and the coupling constant $\beta$ produce relatively small changes to the potential barrier compared with the magnetic charge $Q_m$. Therefore, we only present the relative difference $\Delta V_{\rm eff}$ in the lower panels of Fig.~\ref{fig:effective_potential}.

\begin{figure} 
	\centering 
	
	\begin{minipage}[b]{0.45\textwidth} 
		\centering
		\includegraphics[width=\linewidth, height=4.25cm]{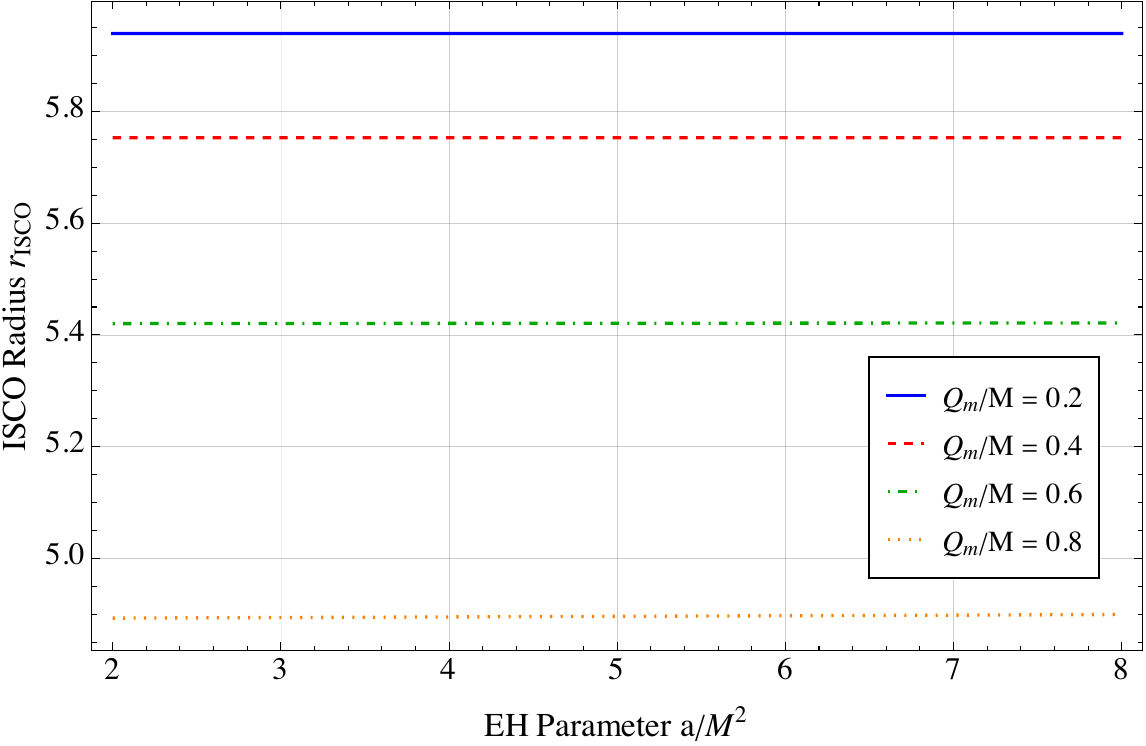} 
	\end{minipage}
	%\hfill 
	\begin{minipage}[b]{0.45\textwidth} 
		\centering
		\includegraphics[width=\linewidth, height=4.25cm]{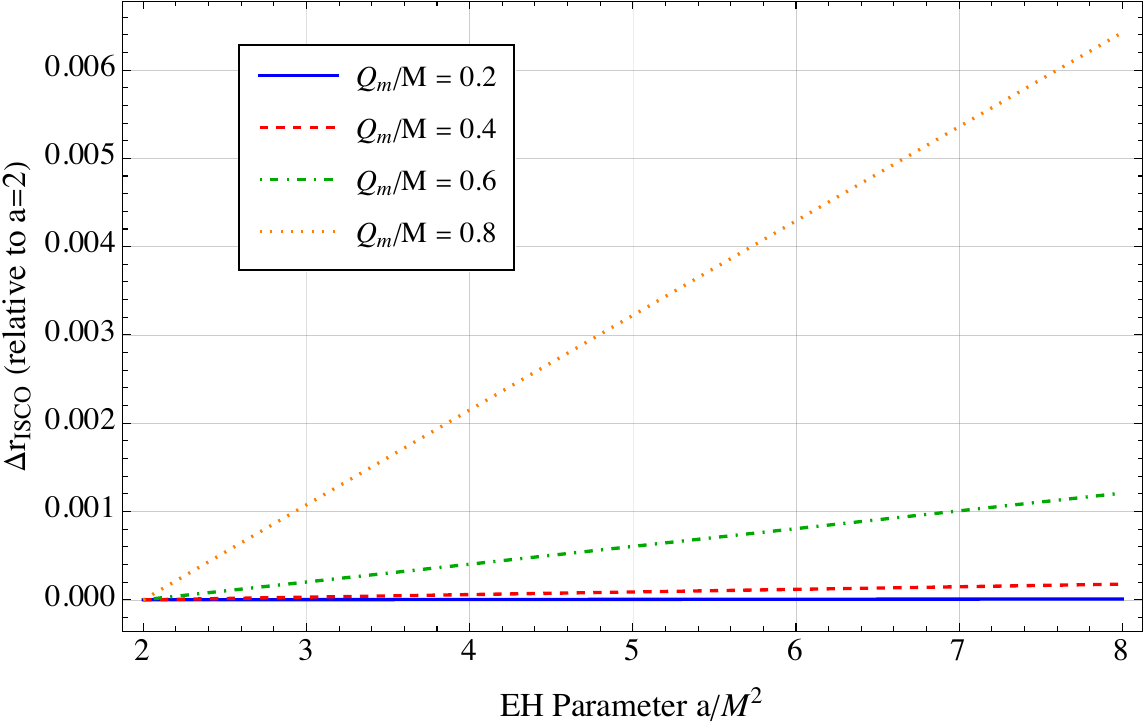} 
	\end{minipage}
	
	\vspace{0.3cm}

	\begin{minipage}[b]{0.45\textwidth} 
		\centering
		\includegraphics[width=\linewidth, height=4.25cm]{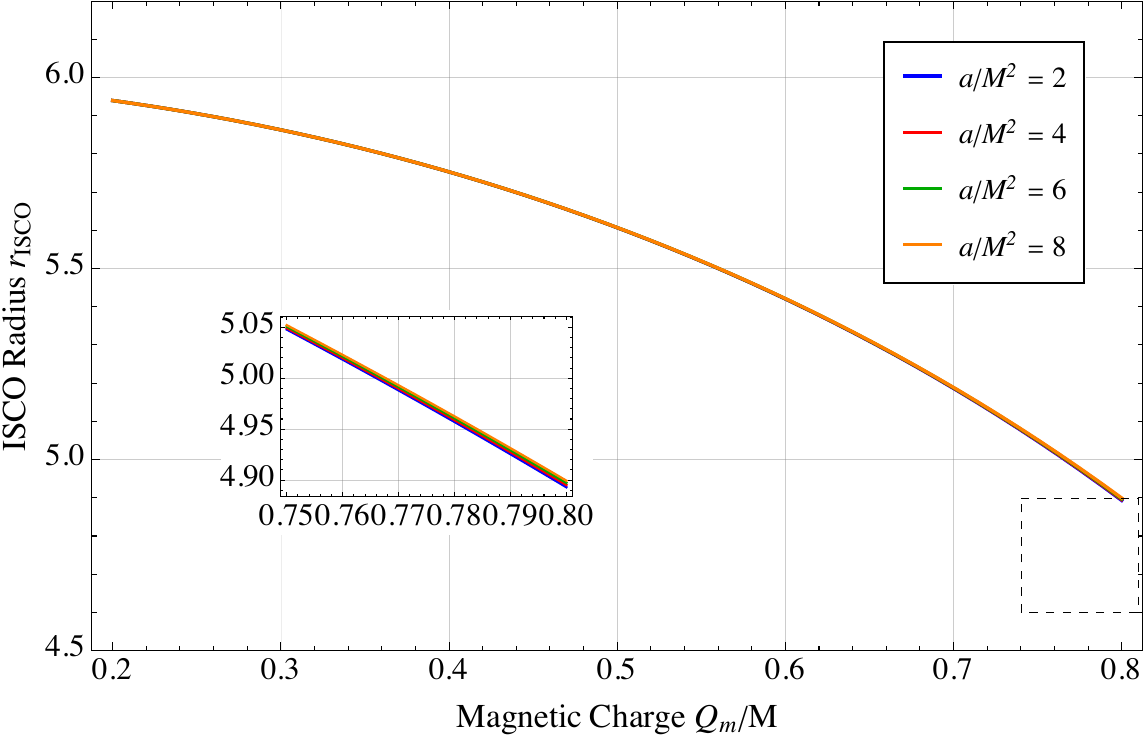} 
	\end{minipage}
	
	\caption{The ISCO radius for the magnetically charged black hole in $f(R,T)$ gravity coupled with Euler-Heisenberg theory. Upper left panel: the ISCO radius $r_{\mathrm{ISCO}}$ as a function of the Euler-Heisenberg parameter $a$. Upper right panel: the variation of the ISCO radius relative to the reference case $a_{\text{ref}}/M^2=2$, namely $\Delta r_{\mathrm{ISCO}} = r_{\mathrm{ISCO}}(a)-r_{\mathrm{ISCO}}(a_{\text{ref}})$, to clearly illustrate the variational trend affected by nonlinear electrodynamics under different magnetic charges $Q_m$. Lower panel: dependence of the ISCO radius $r_{\mathrm{ISCO}}$ on the magnetic charge $Q_m$ for different values of $a$. In all panels, the coupling strength is fixed at $\beta=0.4$.} 
	\label{fig:isco_combined} 
\end{figure}

\begin{figure}
	\centering 
	
	\vspace{0.2cm}
	
	\begin{minipage}[b]{0.45\textwidth} 
		\centering
		\includegraphics[width=\linewidth, height=4.25cm]{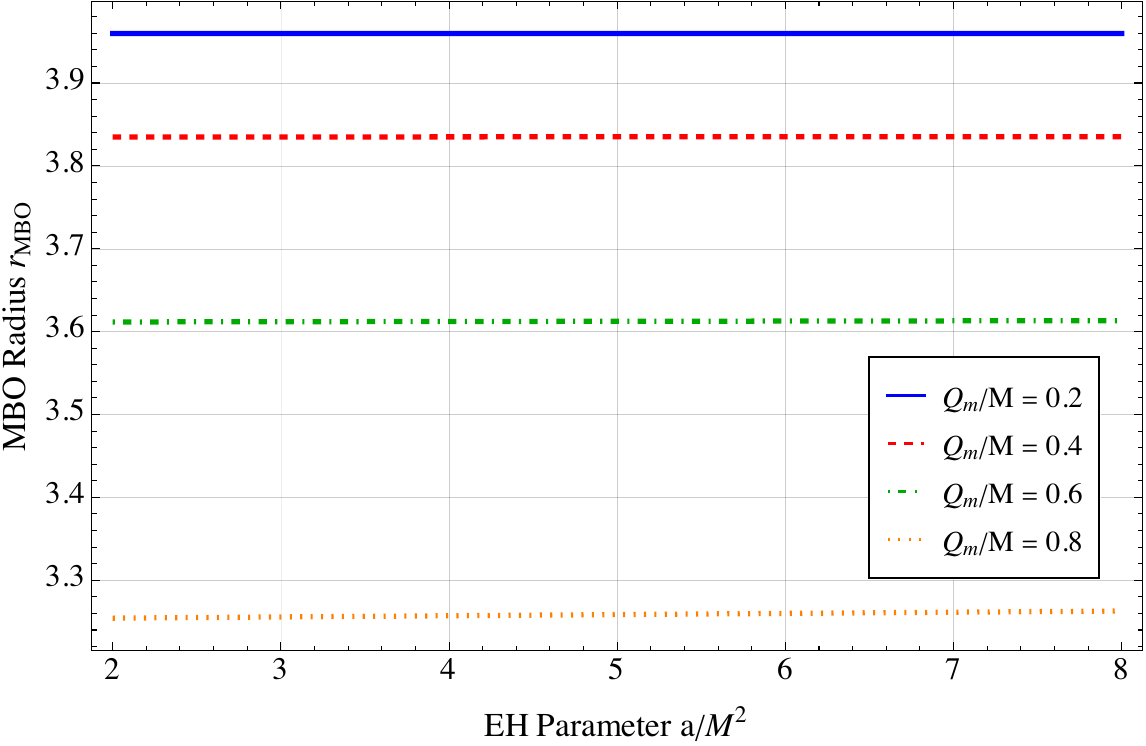} 
	\end{minipage}
	%\hfill 
	\begin{minipage}[b]{0.45\textwidth} 
		\centering
		\includegraphics[width=\linewidth, height=4.25cm]{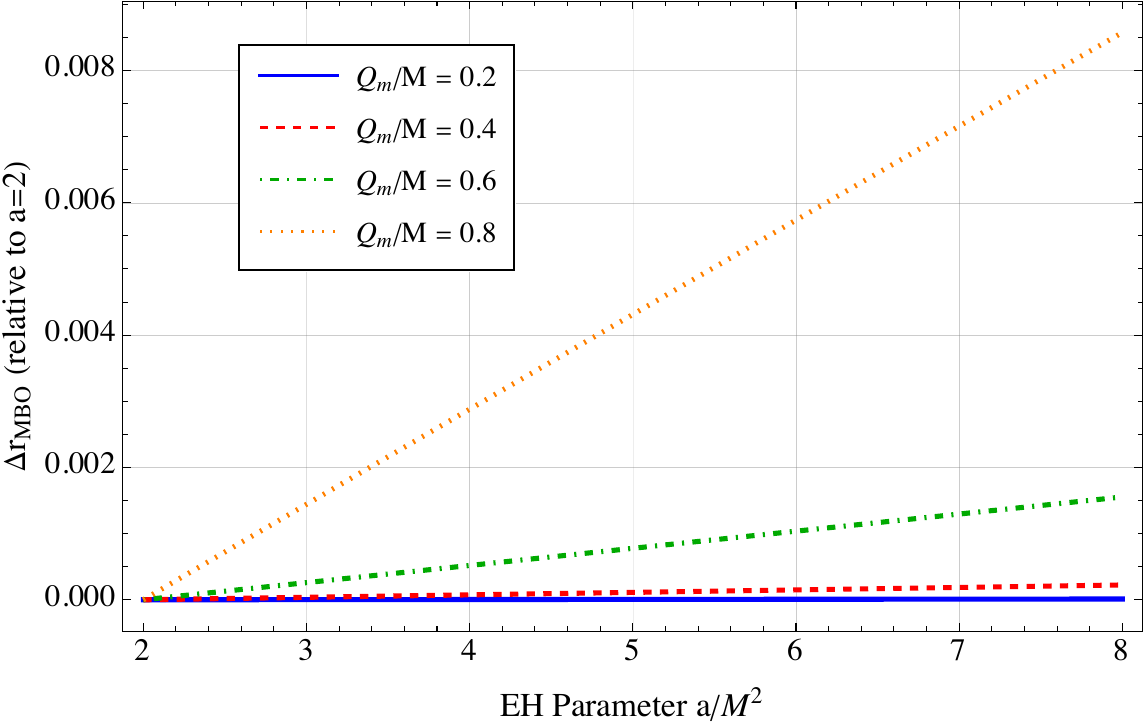} 
	\end{minipage}
	
	\vspace{0.3cm}

	\begin{minipage}[b]{0.45\textwidth} 
		\centering
		\includegraphics[width=\linewidth, height=4.25cm]{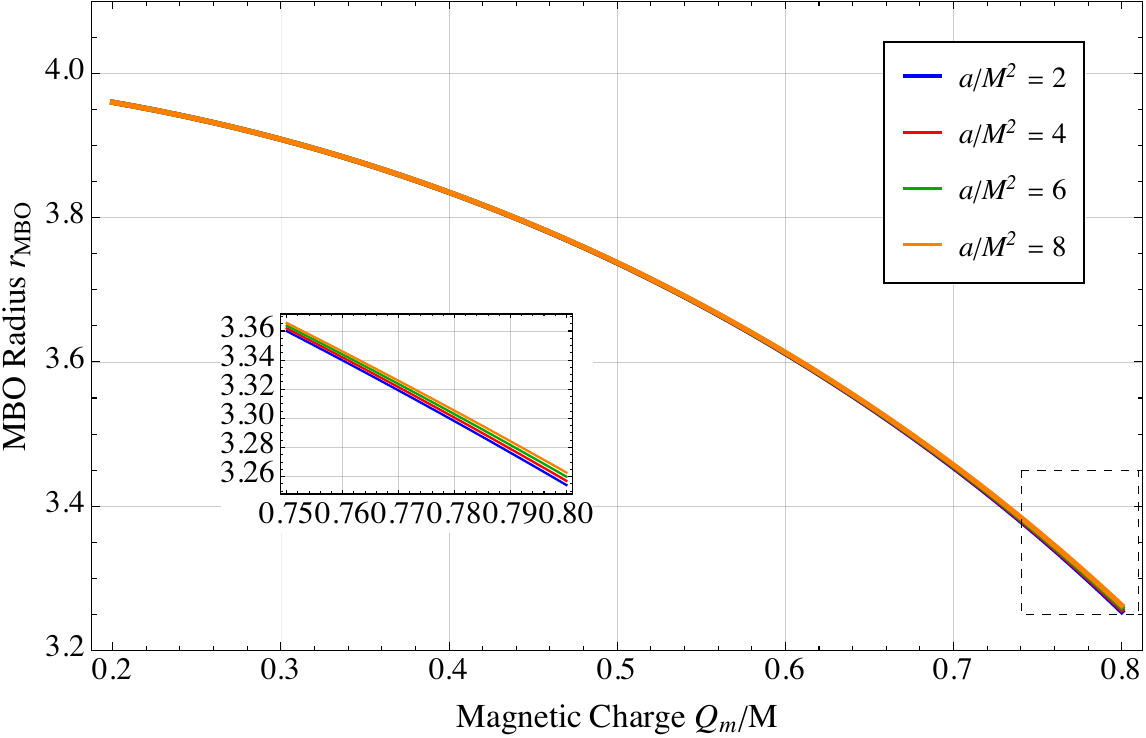} 
	\end{minipage}
	
	\caption{The MBO radius for the magnetically charged black hole in $f(R,T)$ gravity coupled with Euler-Heisenberg theory. Upper left panel: the MBO radius $r_{\mathrm{MBO}}$ as a function of the Euler-Heisenberg parameter $a$. Upper right panel: the variation of the MBO radius relative to the reference case $a_{\text{ref}}/M^2=2$, namely $\Delta r_{\mathrm{MBO}} = r_{\mathrm{MBO}}(a)-r_{\mathrm{MBO}}(a_{\text{ref}})$, highlighting the subtle differences in the MBO affected by nonlinear electrodynamics under various magnetic charges $Q_m$. Lower panel: dependence of the MBO radius $r_{\mathrm{MBO}}$ on the magnetic charge $Q_m$ for different values of $a$. In all panels, the coupling strength is fixed at $\beta=0.4$.} 
	\label{fig:mbo_combined}
\end{figure}
%The left panels of Figs.~\ref{fig:isco_combined} and \ref{fig:mbo_combined} show the variations of the characteristic radii $\Delta r_{\mathrm{ISCO}}$ and $\Delta r_{\mathrm{MBO}}$ relative to those obtained in the reference case $a=2$, as functions of the Euler-Heisenberg parameter $a$ for different values of the magnetic charge $Q_m$. Meanwhile, the right panels of Figs.~\ref{fig:isco_combined} and \ref{fig:mbo_combined} display the dependence of $r_{\mathrm{ISCO}}$ and $r_{\mathrm{MBO}}$ on the magnetic charge $Q_m$ for different values of $a$.

For this paper, we consider bound and stable orbits of particle motion, in which the equation of effective potential $E^2-V_{\text{eff}}(r)=0$ exhibits two zeros near the turning point $\dot{r}=0$, and the corresponding radii $r=r_a$ and $r=r_p$ represent the apocenter and pericenter distances of the bound orbit. Particularly, the innermost stable orbit (ISCO) and marginally bound orbit (MBO) present extremely significant values for constraining the parameter range ($E$,$L$) for bound orbits in the vicinity of black holes.
For particles and small compact objects moving in stable orbits, the following conditions must be satisfied
\begin{eqnarray}\label{conditions}
	L_{\text{ISCO}} \leq L \quad \text{and} \quad E_{\text{ISCO}} \leq E \leq E_{\text{MBO}} = 1.
\end{eqnarray}
Here, $E_{\text{MBO}}$ is the energy of the particle moving along the marginally bound orbit (MBO), $L_{\text{ISCO}}$ and $E_{\text{ISCO}}$ are the orbital angular momentum and energy of the particle moving along the innermost stable circular orbit (ISCO), respectively. The MBO is the circular bound orbit with the minimum radius, and its energy satisfies $E_{\text{MBO}}=V_{\text{eff}} = 1$. Given the effective potential Eq.~(\ref{Effective potential}) for the particle's motion, the MBO satisfies the following conditions:
\begin{equation}\label{condition one}
E_\text{MBO}=V_{\text{eff}} = 1, \quad \frac{dV_{\text{eff}}}{dr} = 0.
\end{equation}
The innermost stable circular orbit (ISCO) corresponds to the smallest radius at which a test particle can remain in a stable circular orbit around the massive black hole, marking the inner edge of the accretion disk and the transition to plunging motion, whose conditions are given by:
\begin{equation}\label{condition two}
	\dot{r} = 0, \quad \frac{dV_{\text{eff}}}{dr} = 0, \quad \frac{d^{2}V_{\text{eff}}}{dr^{2}} = 0.
\end{equation}

Naturally, the characteristic radii $r_{\mathrm{MBO}}$ and $r_{\mathrm{ISCO}}$ are also completely determined by black hole parameters $Q_m$, $a$, and $\beta$ appearing in the metric function $f(r)$. 
By numerically solving the orbital conditions for ISCO and MBO given in Eqs.~(\ref{condition one}) and (\ref{condition two}), Figs.~\ref{fig:isco_combined} and \ref{fig:mbo_combined} illustrate the dependence of $r_{\mathrm{ISCO}}$ and $r_{\mathrm{MBO}}$ on the Euler-Heisenberg parameter $a$ and the magnetic charge $Q_m$. 
We can observe that both $r_{\mathrm{MBO}}$ and $r_{\mathrm{ISCO}}$ decrease as $Q_m$ increases, while they increase slowly with increasing $a$. To more clearly display the influence of the nonlinear electromagnetic effect on the ISCO and MBO radii, the upper right panels of Figs.~\ref{fig:isco_combined} and \ref{fig:mbo_combined} show the variations $\Delta r_{\mathrm{ISCO}}$ and $\Delta r_{\mathrm{MBO}}$ (relative to the reference case $a_{\text{ref}}/M^2=2$) as functions of $a$ for different values of $Q_m$, whereas the upper left panels display the dependence of $r_{\mathrm{ISCO}}$ and $r_{\mathrm{MBO}}$ on $Q_m$ for different values of $a$.

\section{Periodic Orbits}\label{sec3}

In this section, we focus on the periodic orbits around a magnetically charged black hole in $f(R, T)$ gravity coupled
with Euler-Heisenberg electrodynamics. A key physical quantity characterizing the periodic orbits is the precession
parameter $q$, which takes rational values for closed periodic orbits. The periodic orbits can be characterized by a
series of configurations $(z, w, v)$, with the corresponding precession parameter given by $q = w + \frac{v}{z}$. In Subsection \ref{sec3a}, we discuss the properties of the precession parameter $q$. In Subsection \ref{sec3b}, we investigate the periodic-orbit trajectories corresponding to different configurations $(z, w, v)$.

\subsection{Precession Parameter} \label{sec3a}

In strong gravitational fields, bound orbits of test particles (or small compact objects) typically undergo orbital precession, whose magnitude is described by a precession parameter $q$. For periodic orbits, this parameter plays a particularly important role, since the periodicity of orbits imposes a strong constraint that restricts the precession parameter to rational values.

Following the taxonomy of periodic orbits proposed in
ref.~\cite{Levin:2008mq}, a closed periodic orbit can be characterized by three integers $(z,w,v)$. Here, $z$ denotes the number of zooms or distinct leaves, $w$ denotes the number of additional whirls near periapsis, and $v$ describes the vertex jump between successive apoapses. The characteristic precession parameter $q$ associated with a periodic configuration $(z,w,v)$
is given by
\begin{equation}\label{q}
    q=\frac{\Delta\phi_r}{2\pi}-1=w+\frac{v}{z}.
\end{equation}
Here, $\Delta\phi_r$ is the total azimuthal angle accumulated during one complete radial oscillation, from periapsis to apoapsis and back to periapsis. From the equation of motion, it can be expressed as
\begin{align}\label{deltaphi}
    \Delta\phi_r
    &=\oint d\phi
      =2\int_{r_p}^{r_a}\frac{\dot\phi}{\dot r}\,dr \nonumber\\
    &=2\int_{r_p}^{r_a}
      \frac{L\,dr}
      {r^2\sqrt{E^2-f(r)\left(1+\dfrac{L^2}{r^2}\right)}}.
\end{align}
The radii $r_p$ and $r_a$ denote the periapsis and apoapsis, respectively,
and an overdot denotes the differentiation with respect to the proper time
$\tau$.

\begin{figure*}
    \centering
    \begin{minipage}[b]{0.48\textwidth}
        \centering
        \includegraphics[width=\linewidth]{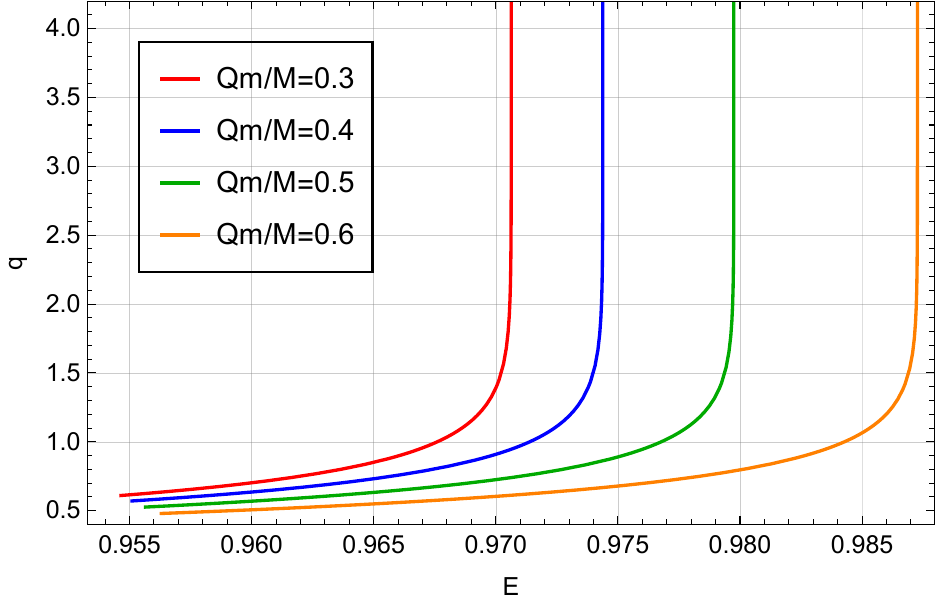}
    \end{minipage}
    \hfill
    \begin{minipage}[b]{0.48\textwidth}
        \centering
        \includegraphics[width=\linewidth]{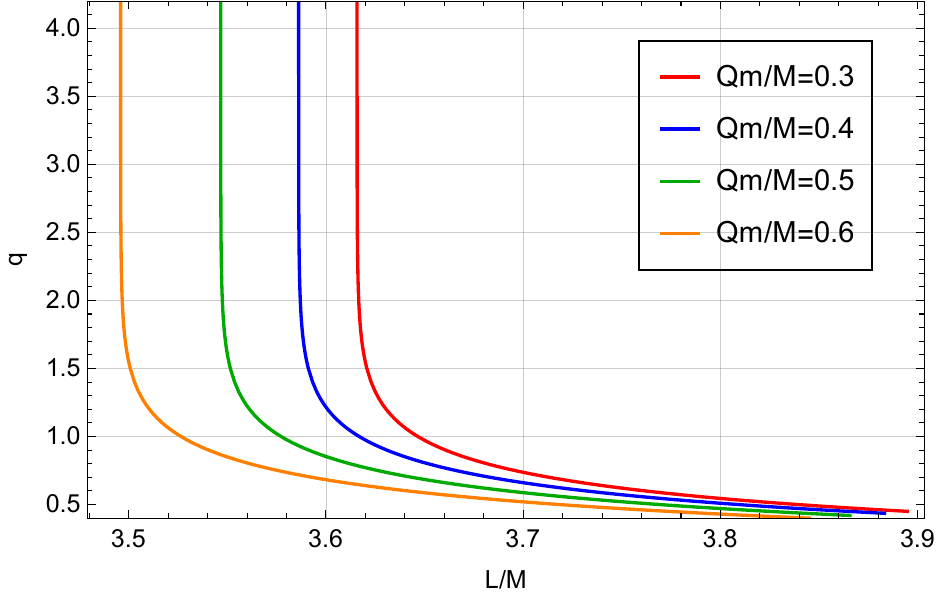}
    \end{minipage}
    \caption{Dependence of the orbital precession parameter $q$ on the reduced energy $E$ (left panel) and angular momentum $L/M$ (right panel). 
    In the left panel, the angular momentum is fixed at
    $L=\frac{1}{2}(L_{\rm MBO}+L_{\rm ISCO})$; in the right panel, the energy is fixed at $E=0.96$. 
    In both panels, we set the coupling strength to $\beta=0.4$,
    and the Euler-Heisenberg parameter is fixed at $a/M^2=5$.}
    \label{fig:q_parameter}
\end{figure*}

Fig.~\ref{fig:q_parameter} shows the dependence of $q$ on the energy and angular momentum of bound orbits. 
For a fixed angular momentum, $q$ increases with $E$ and diverges as the energy approaches its maximal allowed value. 
For a fixed energy, $q$ decreases monotonically with increasing $L$. 
The numerical results also show that $q$ decreases as the magnetic charge $Q_m$ increases under the same orbital energy or angular momentum. 

Interestingly, Fig.~\ref{fig:q_parameter} also indicates that the precession parameter of a bound orbit possesses a minimum value
\begin{equation}
    q_{\min}\leq q<+\infty.
\end{equation}
The lower bound arises when the bound orbit approaches the stable circular orbit at radius $r_0$. Near this radius, the radial oscillation is small and can be described by a linear perturbation $\delta r$. 
For a fixed angular momentum satisfying $L_{\rm ISCO}<L<L_{\rm MBO}$, the allowed energy of a non-circular bound orbit lies in the interval
\begin{equation}
    E_{\min}<E<E_{\max},
\end{equation}
where $E_{\min}$ is the energy of the stable circular orbit at the local minimum of $V_{\rm eff}$, while $E_{\max}$ is the energy of the unstable circular orbit at the potential barrier. 
As $E$ increases from $E_{\min}$ to $E_{\max}$, the trajectory evolves from a nearly circular orbit to a general bound orbit. 
Meanwhile, as $E$ approaches $E_{\max}$, the particle performs an increasing number of revolutions near periapsis when it moves from periapsis to apoapsis. Hence, the accumulated azimuthal angle $\Delta \phi_r$ during one radial period increases with $E$ and diverges as $E\to E_{\max}$. The precession parameter therefore spans the range $q\in[q_{\min},+\infty)$.

To obtain the value of $q_{\text{min}}$ observed in Fig.~\ref{fig:q_parameter}, we analyze the behavior of the particle's motion in the limit of circular orbits. In this limit, the minimum precession parameter is determined
by the ratio of the azimuthal frequency $\omega_\phi$ to the radial epicyclic
frequency $\omega_r$ \cite{Levin:2008ci,Misra:2010pu}:
\begin{align}\label{q_limit}
    q_{\min}
    &=\lim_{r\to r_0}
      \left(\frac{\Delta\phi_r}{2\pi}-1\right) \nonumber\\
    &=\lim_{r\to r_0}
      \left(\frac{T_r\omega_\phi}{T_r\omega_r}-1\right)
      =\lim_{r\to r_0}
      \left(\frac{\omega_\phi}{\omega_r}-1\right),
\end{align}
where $T_r$ is the radial period. For a bound orbit approaching a circular
orbit, the radial perturbation $\delta r$ completes one harmonic oscillation during
$T_r$, and therefore
\begin{equation}
    \omega_r=\frac{2\pi}{T_r}.
\end{equation}
The mean azimuthal frequency is approximated to
\begin{align}
    \omega_\phi
    &=\frac{1}{T_r}\int_0^{T_r}\frac{d\phi}{dt}\,dt
      \simeq \frac{d\phi}{dt},
\end{align}
because $d\phi/dt$ remains nearly constant in the circular-orbit limit.
For a static and spherically symmetric metric, the azimuthal frequency is
obtained from the geodesic equation as
\begin{equation}\label{Omega_phi}
    \omega_\phi^2
    =\left(\frac{d\phi}{dt}\right)^2
    =\left[\frac{L/{r_0}^2}{E/f(r_0)}\right]^2
    =\frac{f'(r_0)}{2r_0},
\end{equation}
where the prime denotes differentiation with respect to $r$.

The radial epicyclic frequency is derived by perturbing the stable circular
orbit according to $r=r_0+\delta r$. Expanding the radial equation to first
order in $\delta r$ gives
\begin{equation}\label{Omega_r_motion}
    \frac{d^2\delta r}{dt^2}+\omega_r^2\delta r=0,
\end{equation}
Using the circular-orbit condition $V_{\rm eff}'(r_0)=0$, the radial frequency can be derived as
\begin{align}\label{Omega_r}
    \omega_r^2
    &=\frac{1}{2\dot t^{\,2}} \cdot
      \left.\frac{d^2V_{\rm eff}}{dr^2}\right|_{r=r_0}
      %\nonumber\\
    =\frac{f(r_0)^2}{2E^2} \cdot V_{\rm eff}''(r_0).
\end{align}
Then the ratio of radial frequency and azimuthal frequency in the limit of circular orbits can be reduced to
\begin{align}\label{Omega_r_explicit}
    \frac{\omega_r^2}{\omega_\phi^2} 
    ={}& \frac{r_0^4}{2L^2} \cdot V_{\rm eff}''(r_0)  \nonumber \\
    ={}& 3f(r_0) -2r_0 f'(r_0) + \frac{r_0 f(r_0) f''(r_0)}{f'(r_0)}
    \nonumber \\
    ={}& 1+\Delta_{\rm correction}.
\end{align}
where we have used the following angular momentum constraint for circular orbits
\begin{equation}
	V_{\rm eff}'(r_0)=0
	\ \ \Leftrightarrow \ \ 
	L^2=\frac{r_0^3 f'(r_0)}{2f(r_0)-r_0 f'(r_0)} .
\end{equation}
Using the explicit form of $V_{\rm eff}$ in Eq.~(\ref{Effective potential}) for the magnetically charged
black-hole spacetime in $f(R,T)$ theory coupled to Euler-Heisenberg electrodynamics, the correction term can be expressed as
\begin{align}\label{Delta_correction}
\Delta_{\rm correction}
={}&-\frac{6M}{r_0}
+\frac{
Q_m^2r_0^9(r_0+3M)
-4Q_m^4r_0^8
+\mathcal{A}\left(15r_0^6-17Mr_0^5-12Q_m^2r_0^4\right)
-24\mathcal{A}^2
}{
 r_0^6\left(Mr_0^5-Q_m^2r_0^4-3\mathcal{A}\right)
},
\end{align}
with
\begin{equation}\label{A_definition}
    \mathcal{A}=\frac{aQ_m^4}{20\pi}(\beta-2\pi).
\end{equation}
It is clearly manifested that the quantity $\Delta_{\rm correction}$ contains the modifications induced by
the magnetic charge and the Euler-Heisenberg parameter. Combining Eqs.~(\ref{q_limit}) and (\ref{Omega_r_explicit}) gives
\begin{equation}\label{q_analytical}
    q_{\min}
    =\lim_{r\to r_0}
    \left(\frac{\omega_\phi}{\omega_r}-1\right)
    =\left(1+\Delta_{\rm correction}\right)^{-1/2}-1.
\end{equation}

In the Newtonian limit $r_0\to\infty$, the correction vanishes and $1+\Delta_{\rm correction}\to1$, so that $q_{\min}\to0$. This reproduces the closed Keplerian ellipse. 
However, in the strong-field region, $1+\Delta_{\rm correction}$ can become substantially smaller than unity, which makes $\omega_r$ smaller than $\omega_\phi$. 
Consequently, a nonzero lower bound $q_\text{min}$ emerges, as shown in Fig.~\ref{fig:q_parameter}.

\subsection{Periodic Orbit Trajectories} \label{sec3b}

For a closed periodic orbit, Eq.~(\ref{q}) restricts $q$ to rational values. 
This requirement imposes a discrete constraint on the orbital energy and angular momentum for each periodic-orbit configuration $(z,w,v)$.
When $L$ is fixed, the allowed energies are obtained from the intersections of the $q$-$E$ curves with the desired rational values $q=w+\frac{v}{z}$, leading to a discrete value of the energy $E$ for each periodic-orbit configuration $(z,w,v)$. Conversely, when $E$ is fixed, the allowed angular momenta are obtained from the intersections of the $q$-$L$ curves with these rational values, and each periodic-orbit configuration $(z,w,v)$ therefore corresponds to a discrete value of the angular momentum $L$.

Once the required energy and angular momentum $(E,L)$ for periodic orbits are obtained, the corresponding orbital trajectories can be calculated by numerically solving the orbital differential equation
\begin{equation}
    \left(\frac{dr}{d\phi}\right)^2
    =
    \frac{r^4}{L^2}
    \left[
        E^2
        -
        f(r)\left(1+\frac{L^2}{r^2}\right)
    \right].
    \label{orbital_equation}
\end{equation}
The orbital trajectories in Cartesian coordinates $(x,y)$ are then obtained through the coordinate transformation
\begin{equation}
    x=r\cos\phi,
    \qquad
    y=r\sin\phi.
    \label{coordinate_transformation}
\end{equation}

\begin{figure*}
	\centering
	\centering
	% 设置子图之间的间距
	\setlength{\tabcolsep}{1pt} 
	
	% --- 第一行 ---
	\begin{minipage}{0.32\textwidth}
		\centering
		% (50,85) 代表 x=50%(水平居中), y=85%(偏上) 的位置
		% \makebox(0,0) 用于确保文字本身在这个坐标点上完美居中
		\begin{overpic}[width=\linewidth]{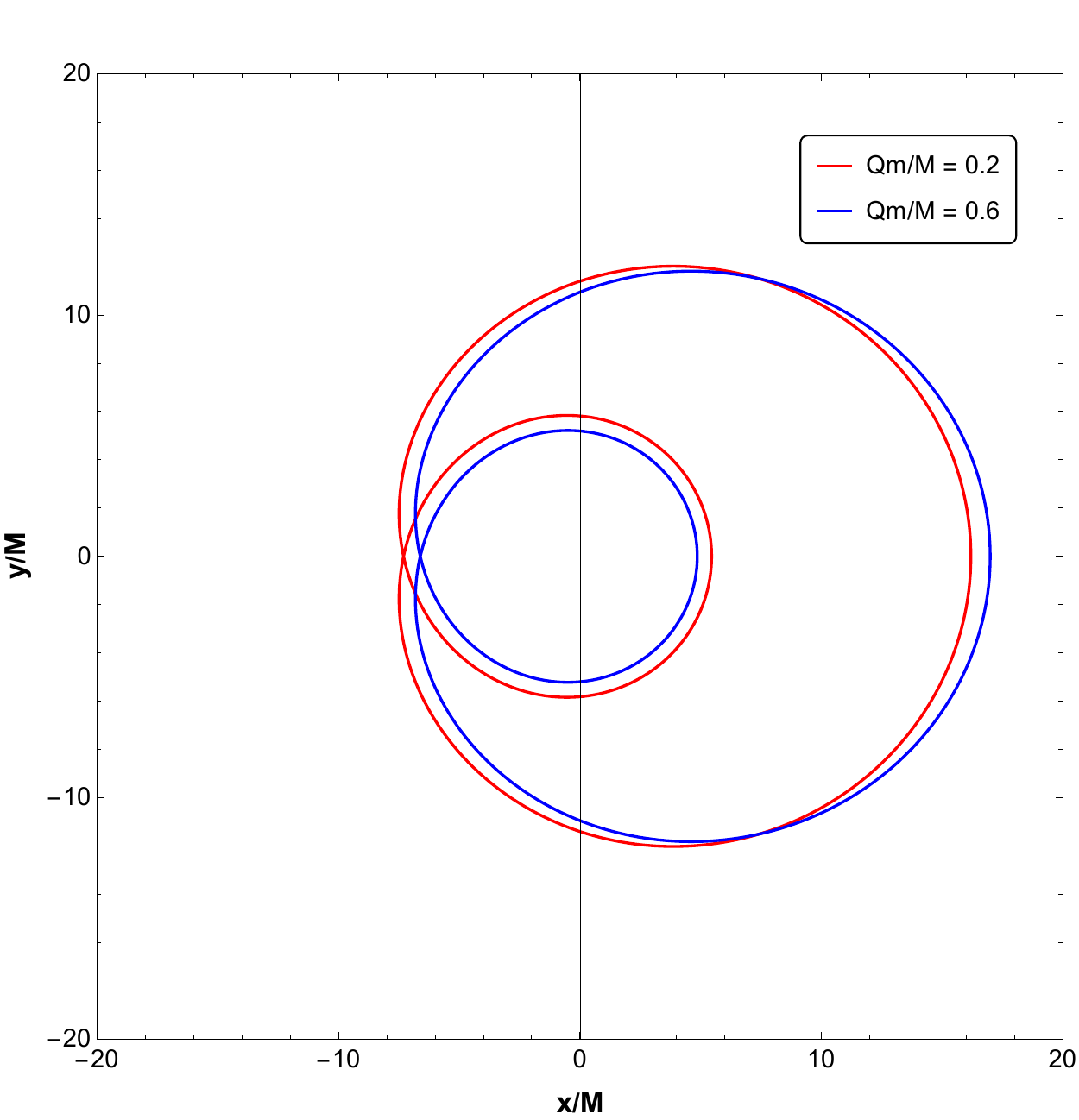}
			\put(50,85){\makebox(0,0){\small $(1, 1, 0)$}} 
		\end{overpic}
	\end{minipage}
	\hfill
	\begin{minipage}{0.32\textwidth}
		\centering
		\begin{overpic}[width=\linewidth]{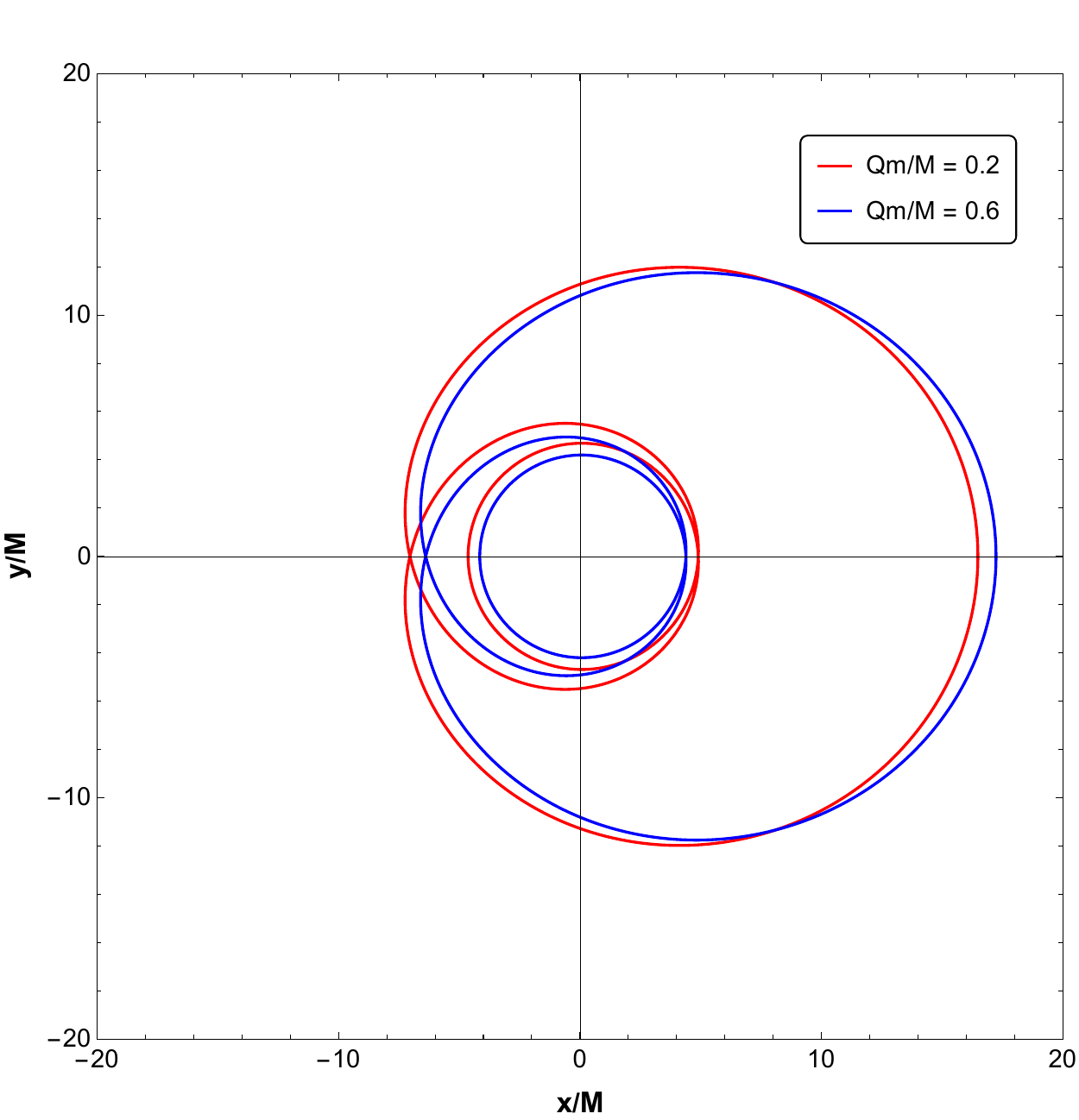}
			\put(50,85){\makebox(0,0){\small $(1, 2, 0)$}}
		\end{overpic}
	\end{minipage}
	\hfill
	\begin{minipage}{0.32\textwidth}
		\centering
		\begin{overpic}[width=\linewidth]{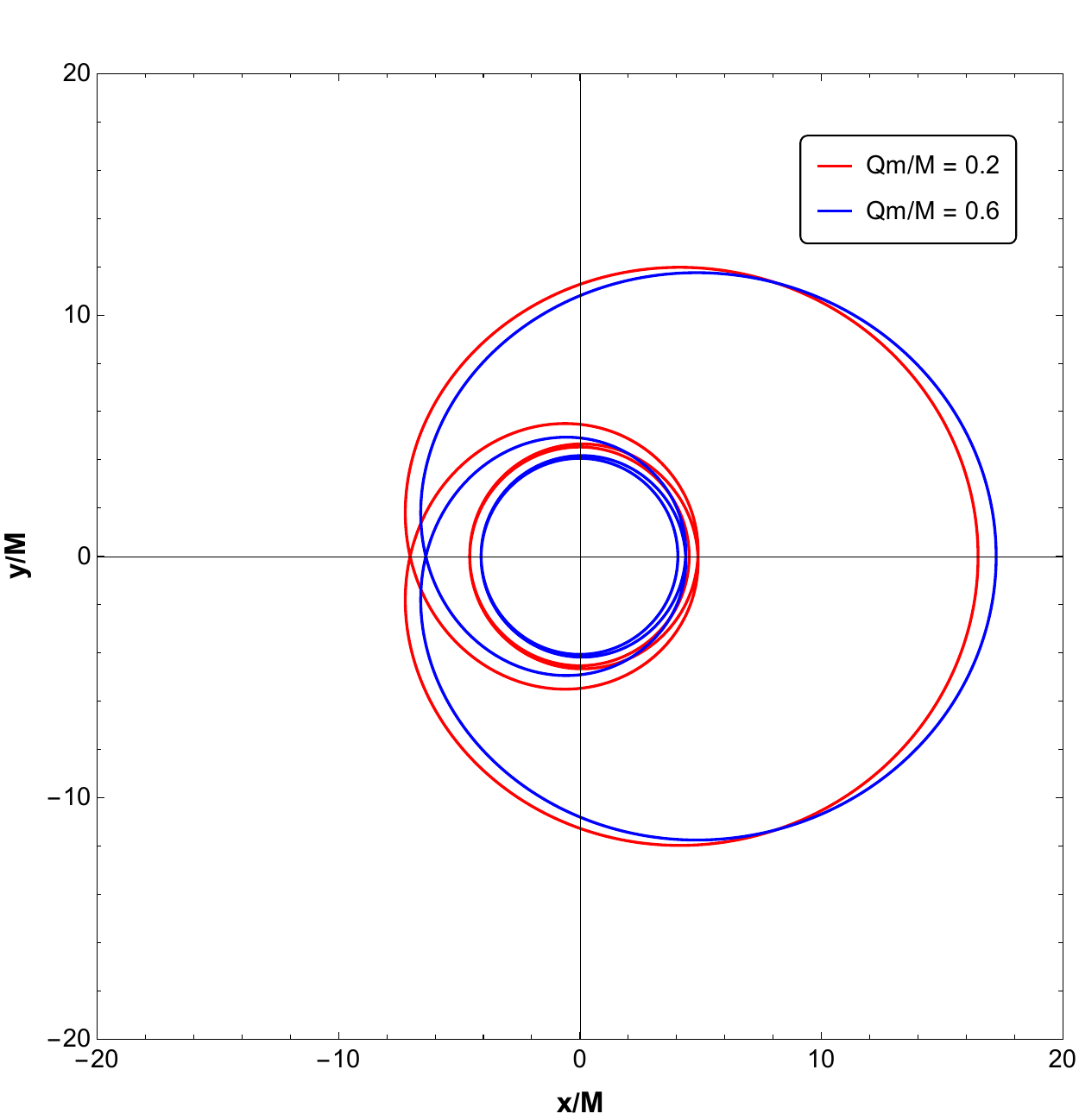}
			\put(50,85){\makebox(0,0){\small $(1, 3, 0)$}}
		\end{overpic}
	\end{minipage}
	\hfill
	\vspace{0.2cm} % 行间距
	
	% --- 第二行 ---
	\begin{minipage}{0.32\textwidth}
		\centering
		\begin{overpic}[width=\linewidth]{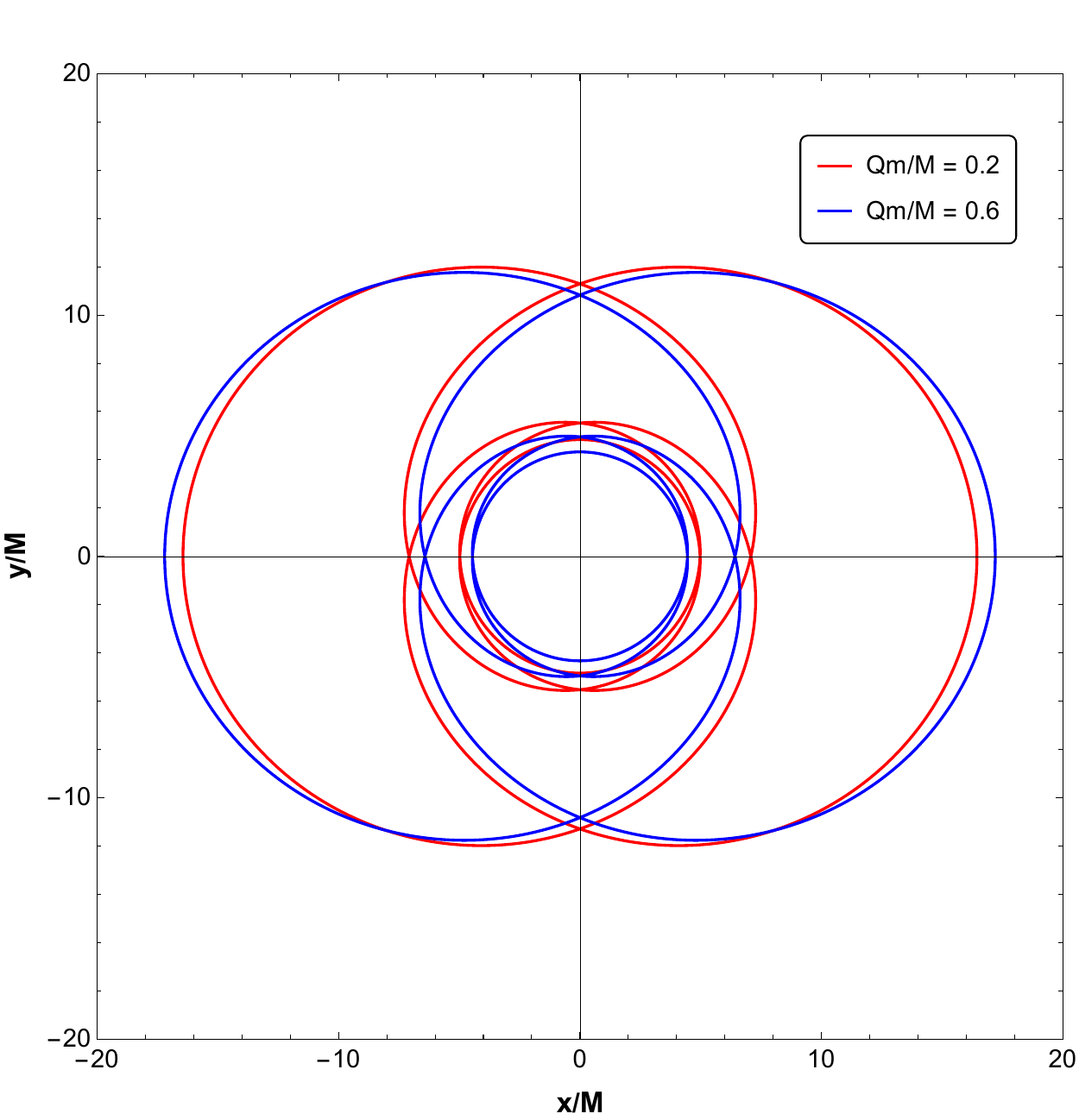}
			\put(50,85){\makebox(0,0){\small $(2, 1, 1)$}}
		\end{overpic}
	\end{minipage}
	\hfill
	\begin{minipage}{0.32\textwidth}
		\centering
		\begin{overpic}[width=\linewidth]{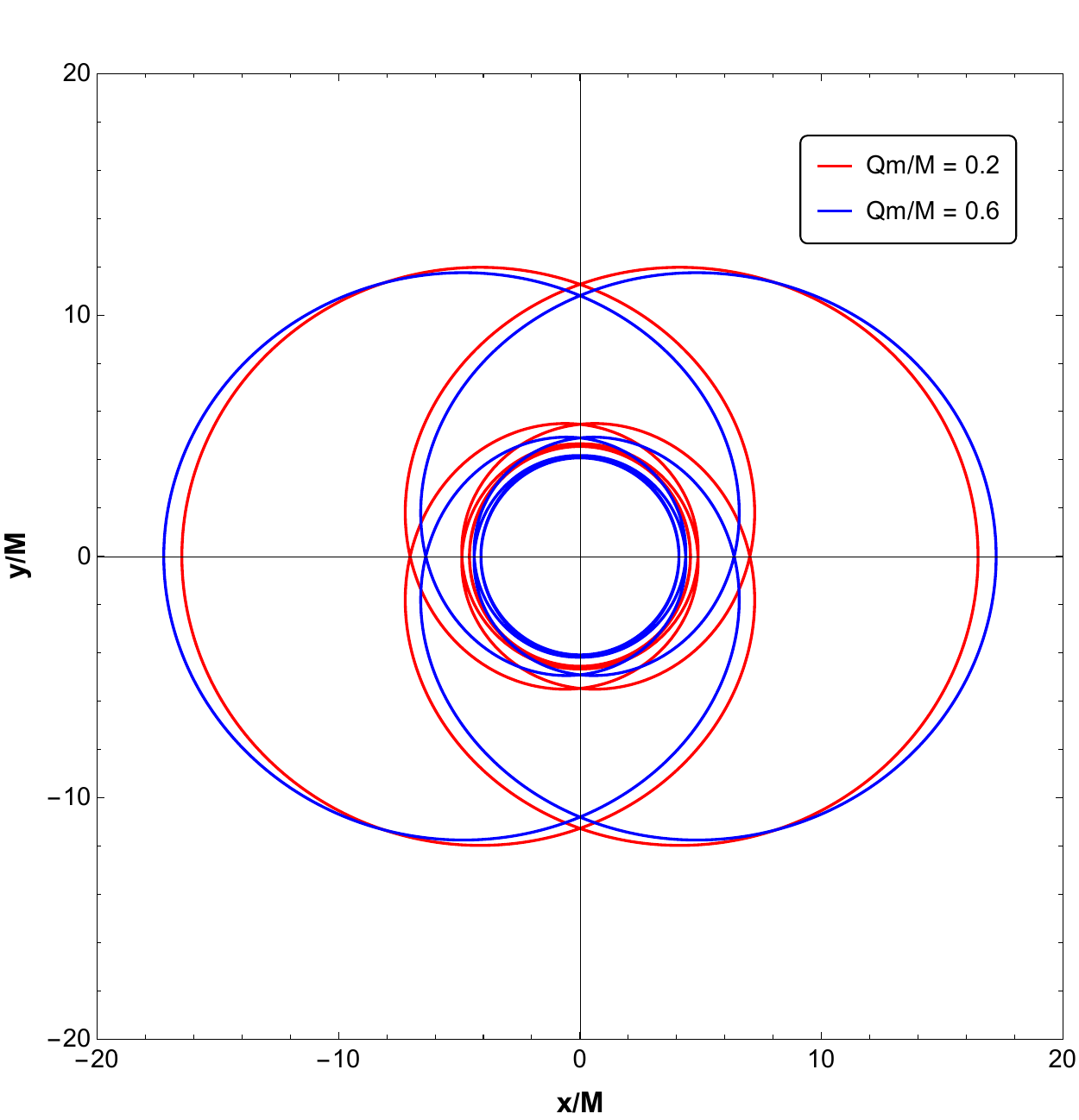}
			\put(50,85){\makebox(0,0){\small $(2, 2, 1)$}}
		\end{overpic}
	\end{minipage}
	\hfill
	\begin{minipage}{0.32\textwidth}
		\centering
		\begin{overpic}[width=\linewidth]{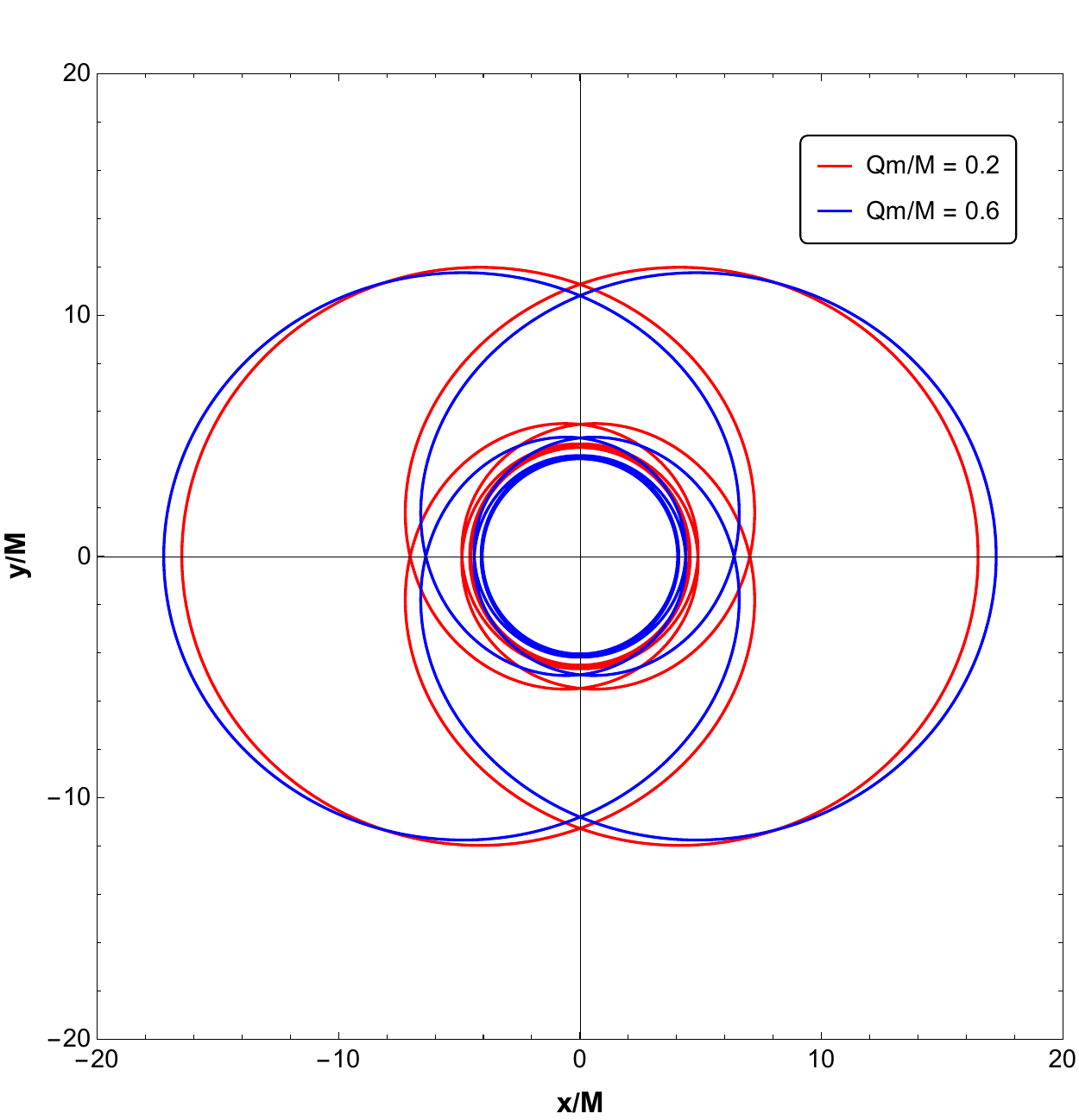}
			\put(50,85){\makebox(0,0){\small $(2, 3, 1)$}}
		\end{overpic}
	\end{minipage}
	\hfill
	\vspace{0.2cm}
	
	\begin{minipage}{0.32\textwidth}
		\centering
		\begin{overpic}[width=\linewidth]{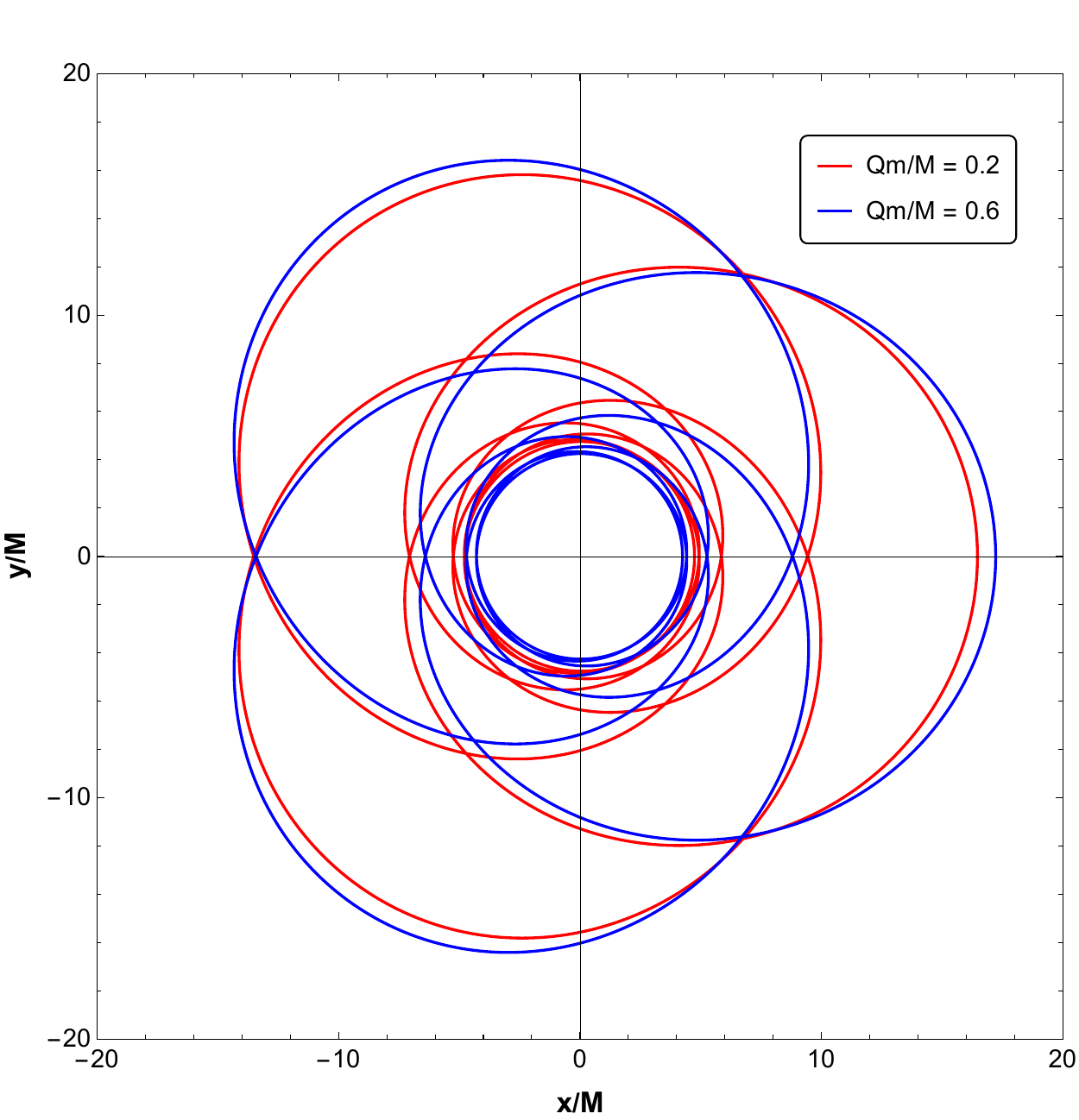}
			\put(50,85){\makebox(0,0){\small $(3, 1, 2)$}}
		\end{overpic}
	\end{minipage}
	\hfill
	\begin{minipage}{0.32\textwidth}
		\centering
		\begin{overpic}[width=\linewidth]{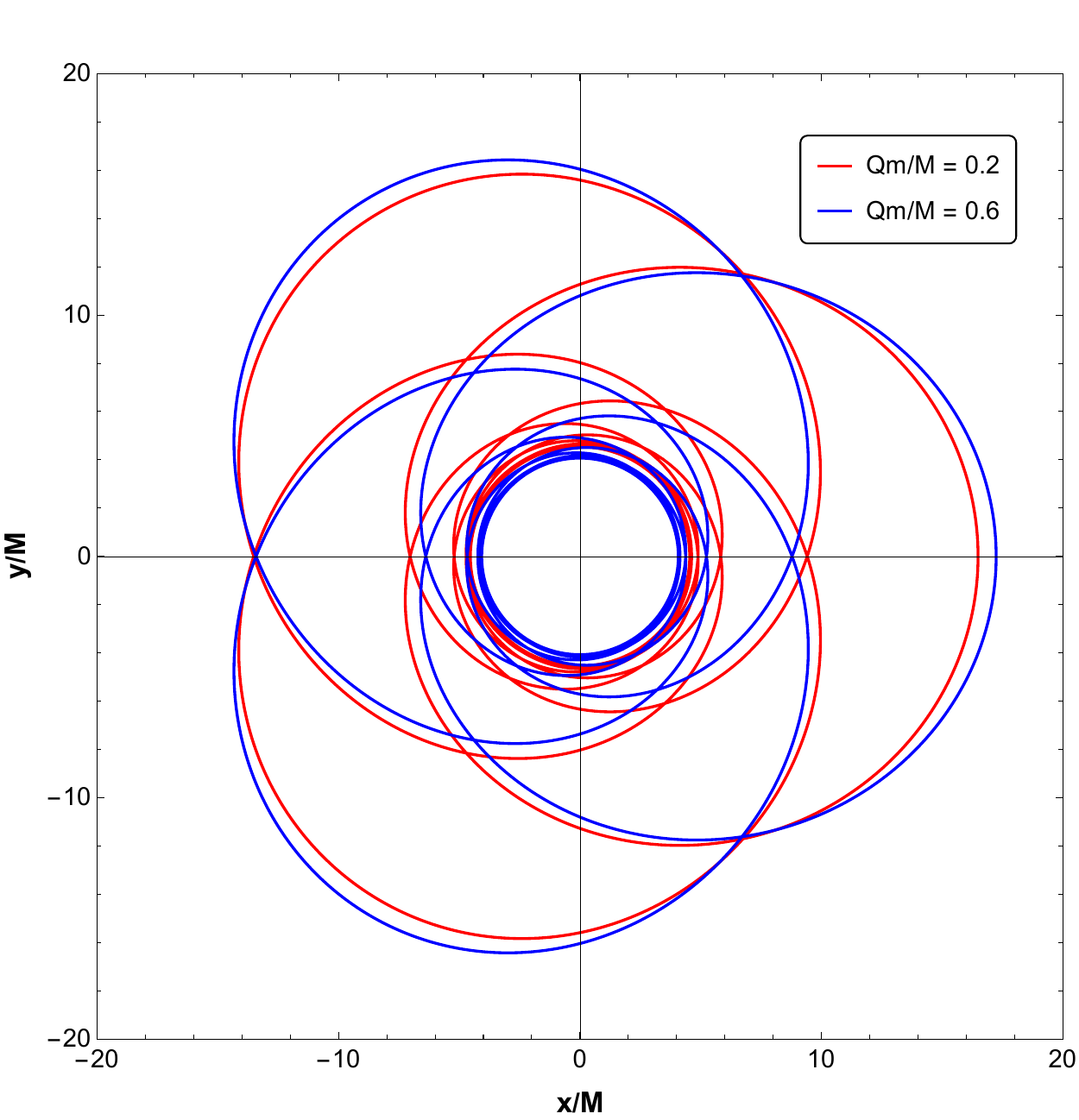}
			\put(50,85){\makebox(0,0){\small $(3, 2, 2)$}}
		\end{overpic}
	\end{minipage}
	\hfill
	\begin{minipage}{0.32\textwidth}
		\centering
		\begin{overpic}[width=\linewidth]{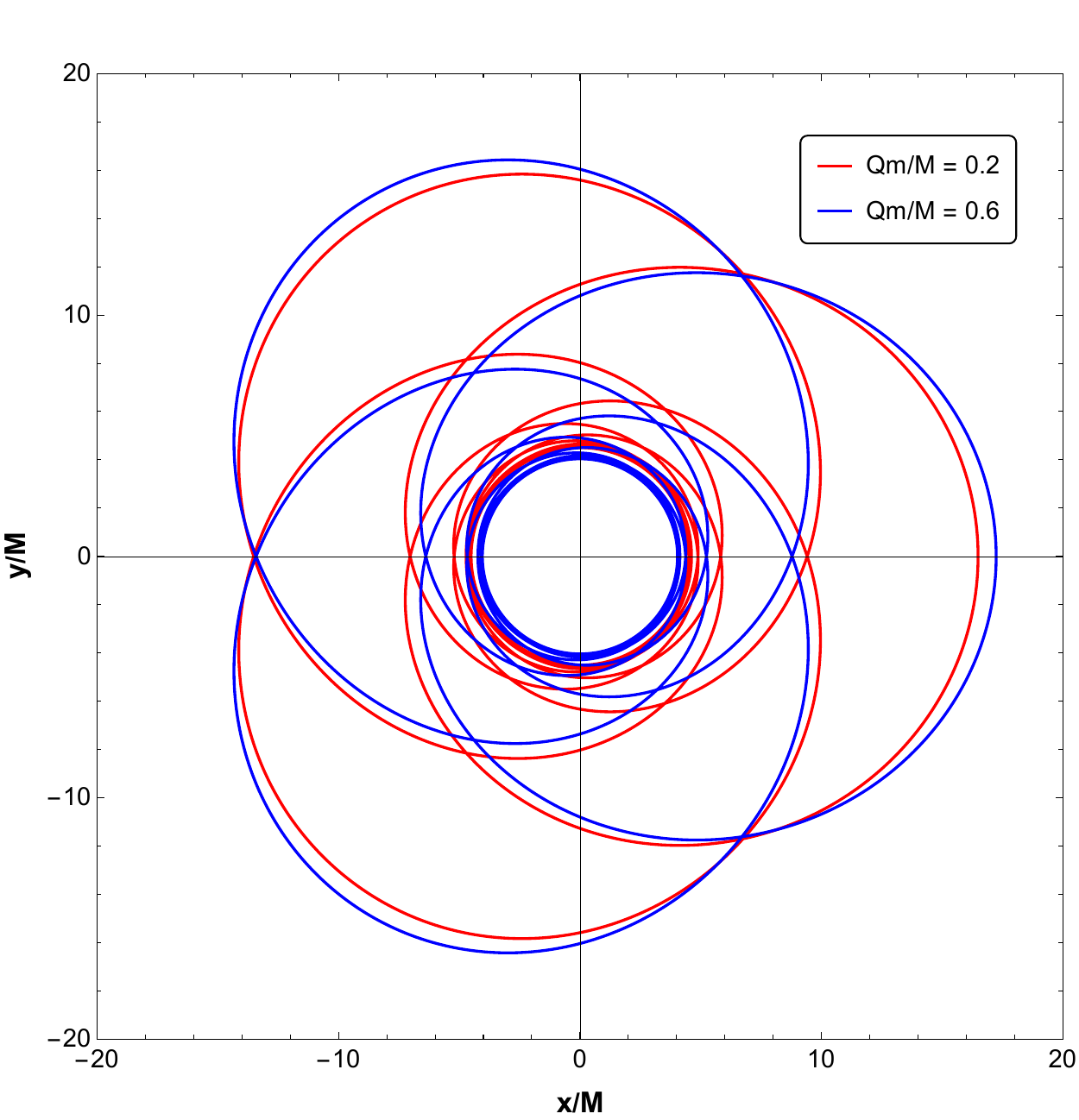}
			\put(50,85){\makebox(0,0){\small $(3, 3, 2)$}}
		\end{overpic}
	\end{minipage}
	\hfill
	\vspace{0.2cm}
	
	\caption{Periodic orbits characterized by different configurations $(z,w,v)$
		for fixed energy $E=0.96$. The nonlinear electromagnetic parameter and coupling strength are fixed at $a/M^2=5$ and $\beta=0.4$. The red image and the blue image respectively represent $Q_m/M=0.2$ and $Q_m/M=0.6$. For each configuration, the angular momenta are selected from Table \ref{tab:angular_momentum_L} to satisfy $q=w+\frac{v}{z}$.}
	\label{fig:orbits_fixedE}
\end{figure*}

In the following, we take $E=0.96$ as a representative value, while qualitatively similar results can be obtained for other admissible energies satisfying the requirement for bound orbits, $E_{\text{ISCO}}<E<E_{\text{MBO}}$. The resulting numerical angular momentum values $L$ for several orbital configurations and magnetic charges are listed in Table~\ref{tab:angular_momentum_L}. These data also confirm the negative correlation between $q$ and $L$ at fixed energy (which haven been illustrated in the right panel of Fig.~\ref{fig:q_parameter}). The trajectories of periodic orbits for several configurations $(z,w,v)$ are displayed in Fig.~\ref{fig:orbits_fixedE}, comparing cases with different magnetic charges $Q_m$ and highlighting the influence of electrodynamics and highlighting the influence of electrodynamics. 
For the same configuration $(z,w,v)$, we can observe that for larger $Q_m$, the apoapsis of its orbit shifts farther away, and the magnitude of whirl in the periodic orbit becomes smaller.

\begin{table*}[b]
	\caption{Angular momentum $L$ of periodic orbits for different magnetic charges $Q_m$, with fixed $E=0.96$, $a/M^2=5$, and $\beta=0.4$.}
	\label{tab:angular_momentum_L}
	\begin{ruledtabular}
		\begin{tabular}{cccccc}
			$(z,w,v)$ & $q$ & $Q_m/M=0.2$ & $Q_m/M=0.4$ & $Q_m/M=0.6$ & $Q_m/M=0.8$ \\
			\hline
			$(1,1,0)$ & $1$    & 3.667303 & 3.616835 & 3.526524 & 3.382987 \\
			$(2,1,1)$ & $1.5$  & 3.641392 & 3.591102 & 3.500808 & 3.356366 \\
			$(3,1,2)$ & $5/3$  & 3.639143 & 3.588880 & 3.498595 & 3.354050 \\
			$(4,1,3)$ & $1.75$ & 3.638434 & 3.588181 & 3.497901 & 3.353320 \\
			$(1,2,0)$ & $2$    & 3.637228 & 3.586994 & 3.496724 & 3.352080 \\
			$(2,2,1)$ & $2.5$  & 3.636529 & 3.586312 & 3.496051 & 3.351365 \\
			$(3,2,2)$ & $8/3$  & 3.636466 & 3.586250 & 3.495990 & 3.351301 \\
			$(4,2,3)$ & $2.75$ & 3.636446 & 3.586231 & 3.495971 & 3.351280 \\
		\end{tabular}
	\end{ruledtabular}
\end{table*}

\begin{figure}[p]
	\centering
	\includegraphics[width=0.975\textwidth]{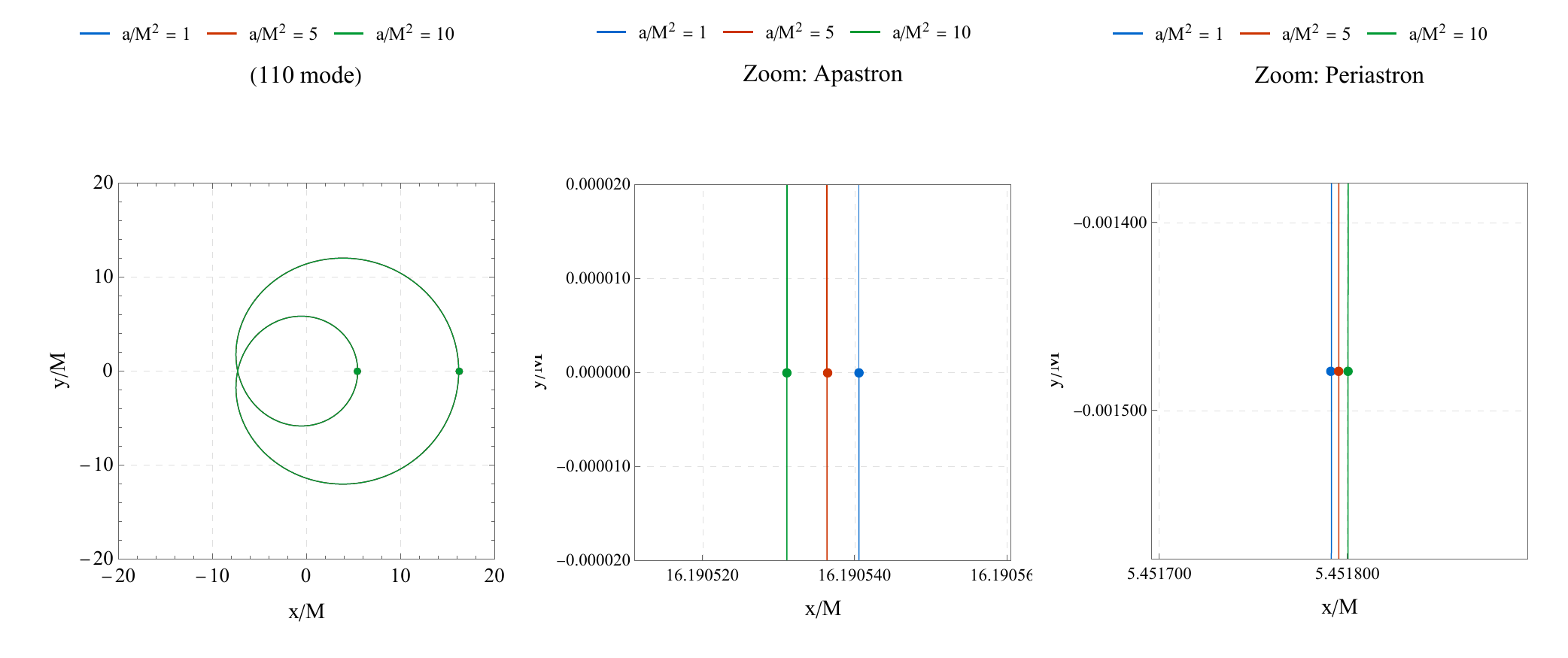}
	
	\vspace{0.3cm}
	
	\includegraphics[width=0.975\textwidth]{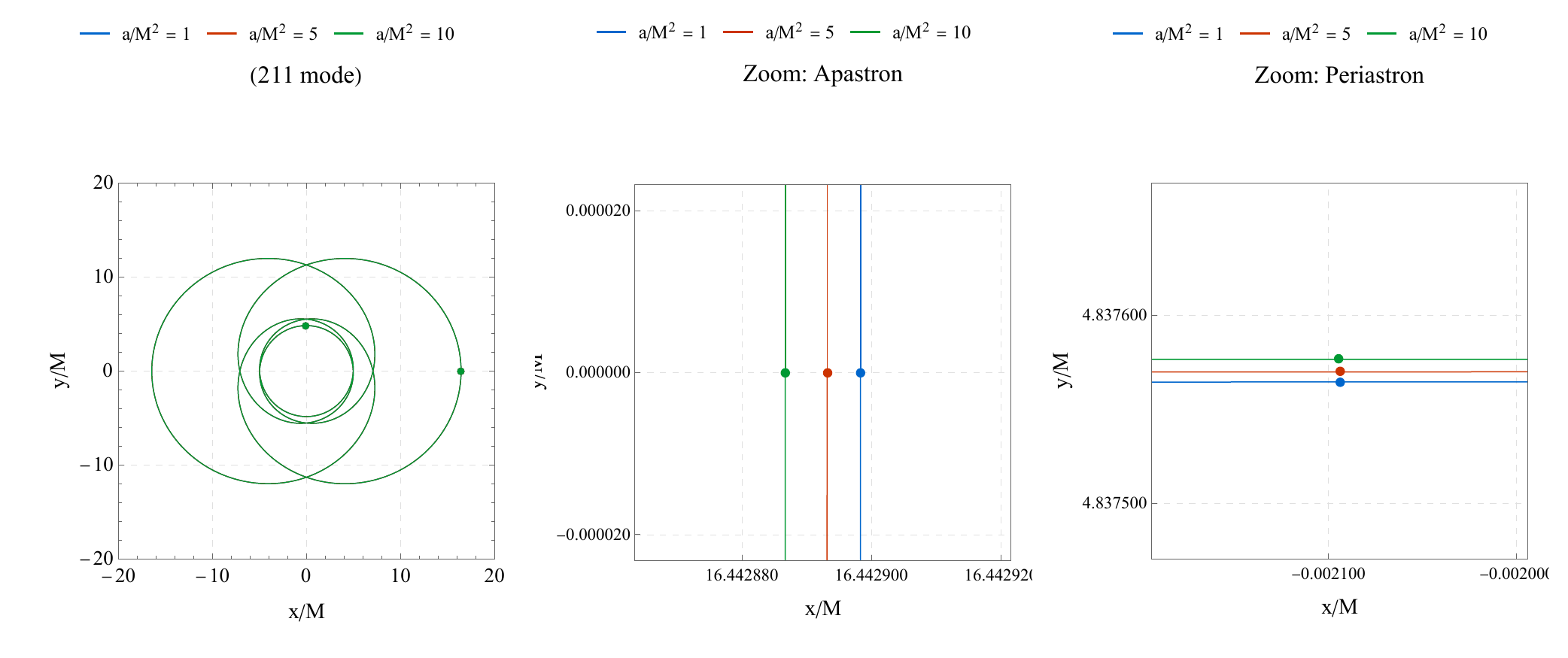}
	\caption{Periodic orbits obtained for different nonlinear electrodynamics parameter values $a$ with fixed energy $E=0.96$, magnetic charge $Q_m/M=0.2$, and coupling parameter $\beta=0.4$. The blue, red, and green curves correspond to $a/M^2=1$, $5$, $10$, respectively. The upper and lower panels show the $(1,1,0)$ and $(2,1,1)$ periodic-orbit configurations, respectively. For each value of $a$, the angular momentum $L$ is determined separately to satisfy the corresponding periodic-orbit condition $q=w+\frac{v}{z}$. 
	The left panel gives the orbital trajectories, and the middle and right panels provide enlarged views of the apastron and periastron positions, respectively.}
	\label{fig:a110}
\end{figure}

Besides the magnetic charge effects, we further investigate how the nonlinear electromagnetic parameter $a$ modifies the periodic-orbit trajectories at fixed energy.
In Fig.~\ref{fig:a110}, we present the periodic orbital trajectories affected by the nonlinear electrodynamics parameter $a$ for the same magnetic charge $Q_m/M=0.2$ and orbital energy $E=0.96$. 
As the nonlinear electrodynamics parameter $a$ increases, the apastron radius is slightly decreased (indicated in the middle panels), while the periastron radius increases slightly (as shown in the right panels). 
Although only the $(2,1,1)$ and $(3,2,2)$ configurations are shown in the figure, this behavior is found to hold for other periodic-orbit configurations. Therefore, compared with the magnetic charge $Q_m$, the nonlinear electrodynamics parameter $a$ has a relatively weaker influence on the periodic-orbit trajectories.

%\begin{minipage}
\begin{table*}[p]
	\caption{Energy $E$ of periodic orbits characterized by $(z,w,v)$ and
		$q$ for different magnetic charges $Q_m$, with fixed $a/M^2=5$,
		$\beta=0.4$, and $L=\frac{1}{2}(L_{\rm ISCO}+L_{\rm MBO})$.}
	\label{tab:energy_q}
	\begin{ruledtabular}
		\begin{tabular}{cccccc}
			$(z,w,v)$ & $q$ & $Q_m/M=0.2$ & $Q_m/M=0.4$ & $Q_m/M=0.6$ & $Q_m/M=0.8$ \\
			\hline
			$(1,1,0)$ & $1$    & 0.9677346 & 0.9640260 & 0.9619330 & 0.9579979 \\
			$(2,1,1)$ & $1.5$  & 0.9679366 & 0.9667979 & 0.9649888 & 0.9616849 \\
			$(3,1,2)$ & $5/3$  & 0.9679978 & 0.9670123 & 0.9652304 & 0.9619902 \\
			$(4,1,3)$ & $1.75$ & 0.9680976 & 0.9670775 & 0.9653045 & 0.9620853 \\
			$(1,2,0)$ & $2$    & 0.9681113 & 0.9671843 & 0.9654268 & 0.9622449 \\
			$(2,2,1)$ & $2.5$  & 0.9681181 & 0.9672412 & 0.9654932 & 0.9622831 \\
			$(3,2,2)$ & $8/3$  & 0.9681560 & 0.9672078 & 0.9654577 & 0.9622885 \\
			$(4,2,3)$ & $2.75$ & 0.9681601 & 0.9672112 & 0.9654612 & 0.9622904 \\
		\end{tabular}
	\end{ruledtabular}
\end{table*}
%\end{minipage}

\begin{figure*}[t]
	\centering
	\centering
	% 设置子图之间的间距
	\setlength{\tabcolsep}{1pt} 
	
	% --- 第一行 ---
	\begin{minipage}{0.32\textwidth}
		\centering
		% (50,85) 代表 x=50%(水平居中), y=85%(偏上) 的位置
		% \makebox(0,0) 用于确保文字本身在这个坐标点上完美居中
		\begin{overpic}[width=\linewidth]{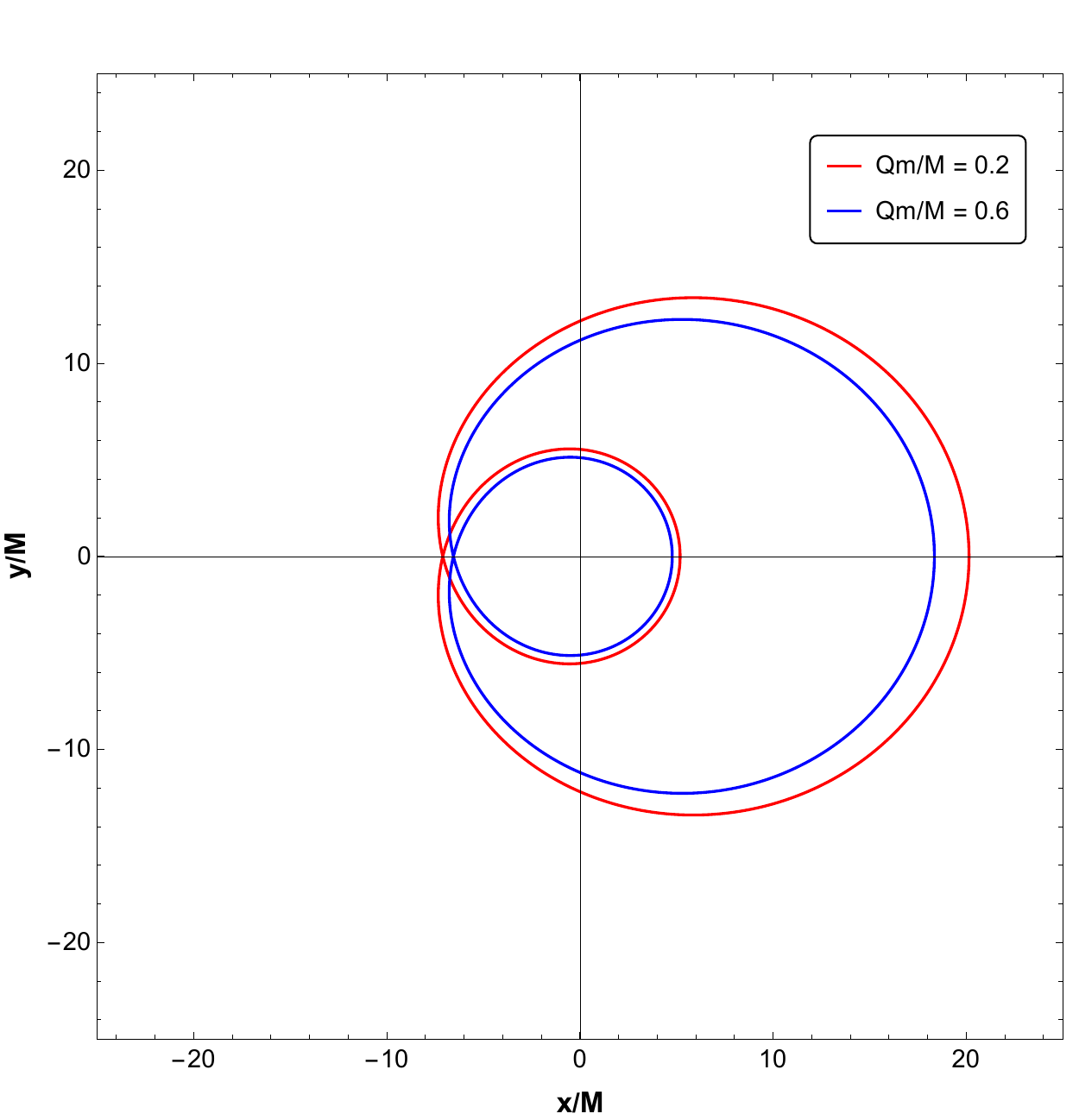}
			\put(50,85){\makebox(0,0){\small $(1, 1, 0)$}} 
		\end{overpic}
	\end{minipage}
	\hfill
	\begin{minipage}{0.32\textwidth}
		\centering
		\begin{overpic}[width=\linewidth]{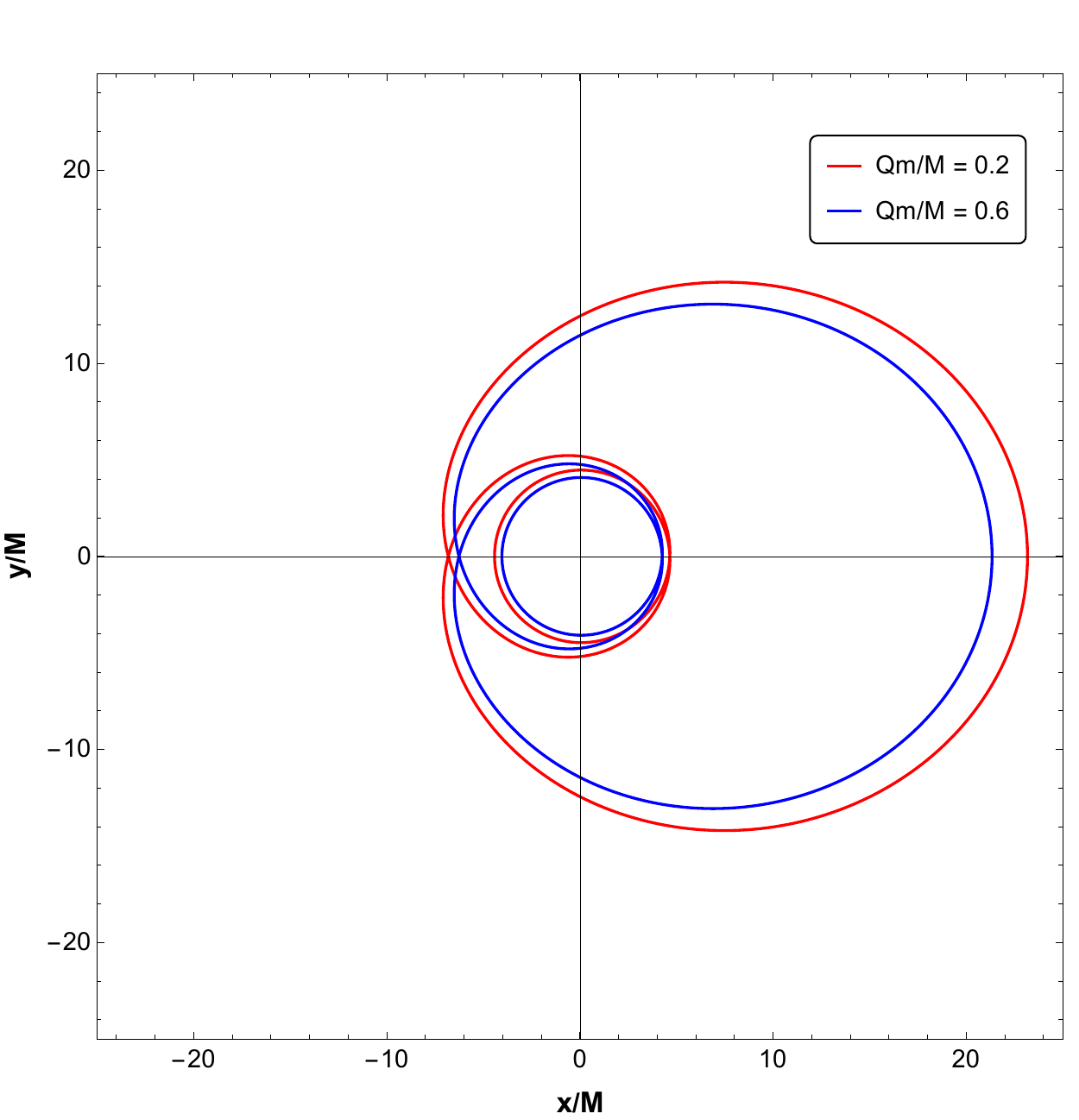}
			\put(50,85){\makebox(0,0){\small $(1, 2, 0)$}}
		\end{overpic}
	\end{minipage}
	\hfill
	\begin{minipage}{0.32\textwidth}
		\centering
		\begin{overpic}[width=\linewidth]{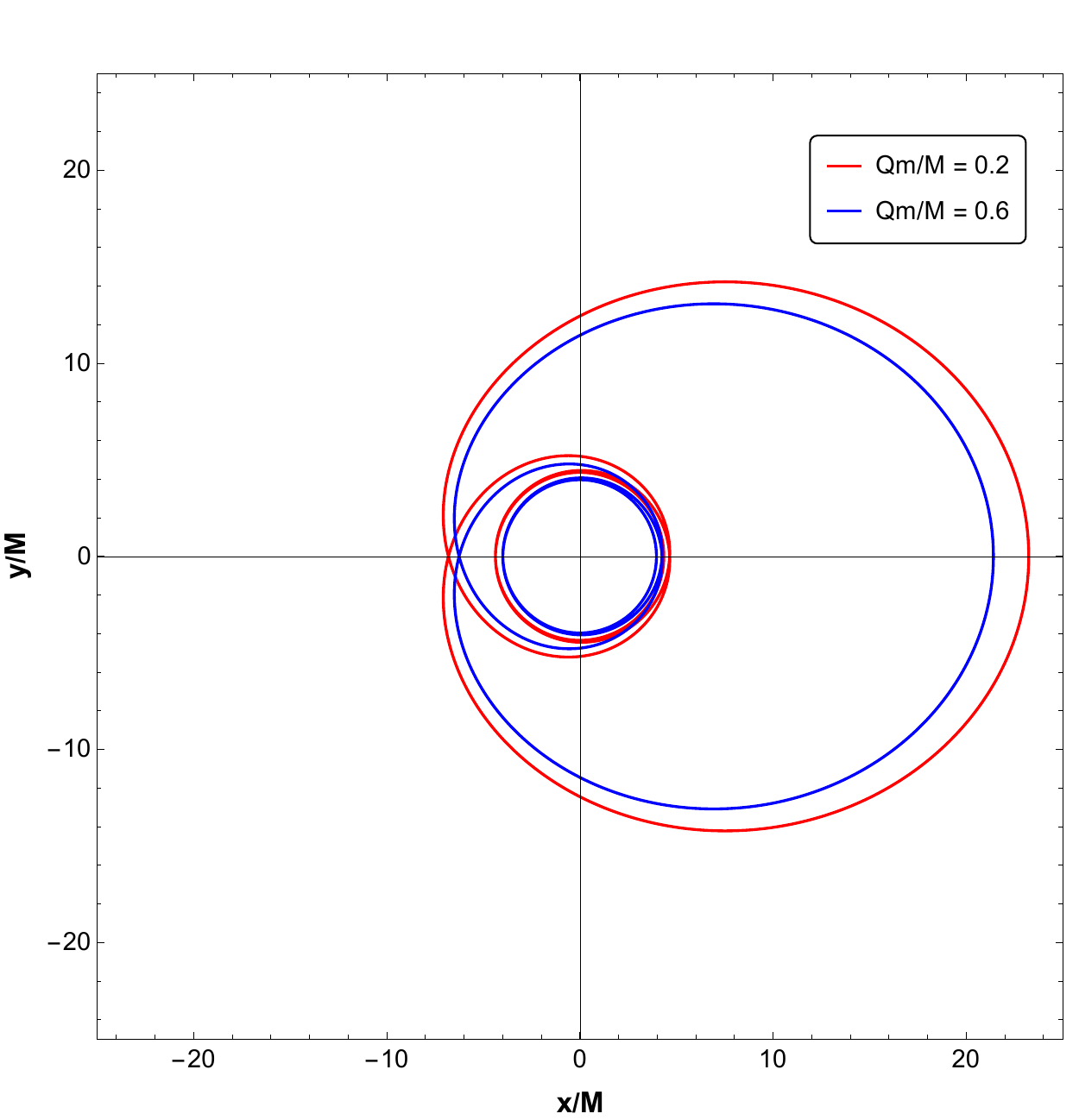}
			\put(50,85){\makebox(0,0){\small $(1, 3, 0)$}}
		\end{overpic}
	\end{minipage}
	\hfill
	\vspace{0.2cm} % 行间距
	
	% --- 第二行 ---
	\begin{minipage}{0.32\textwidth}
		\centering
		\begin{overpic}[width=\linewidth]{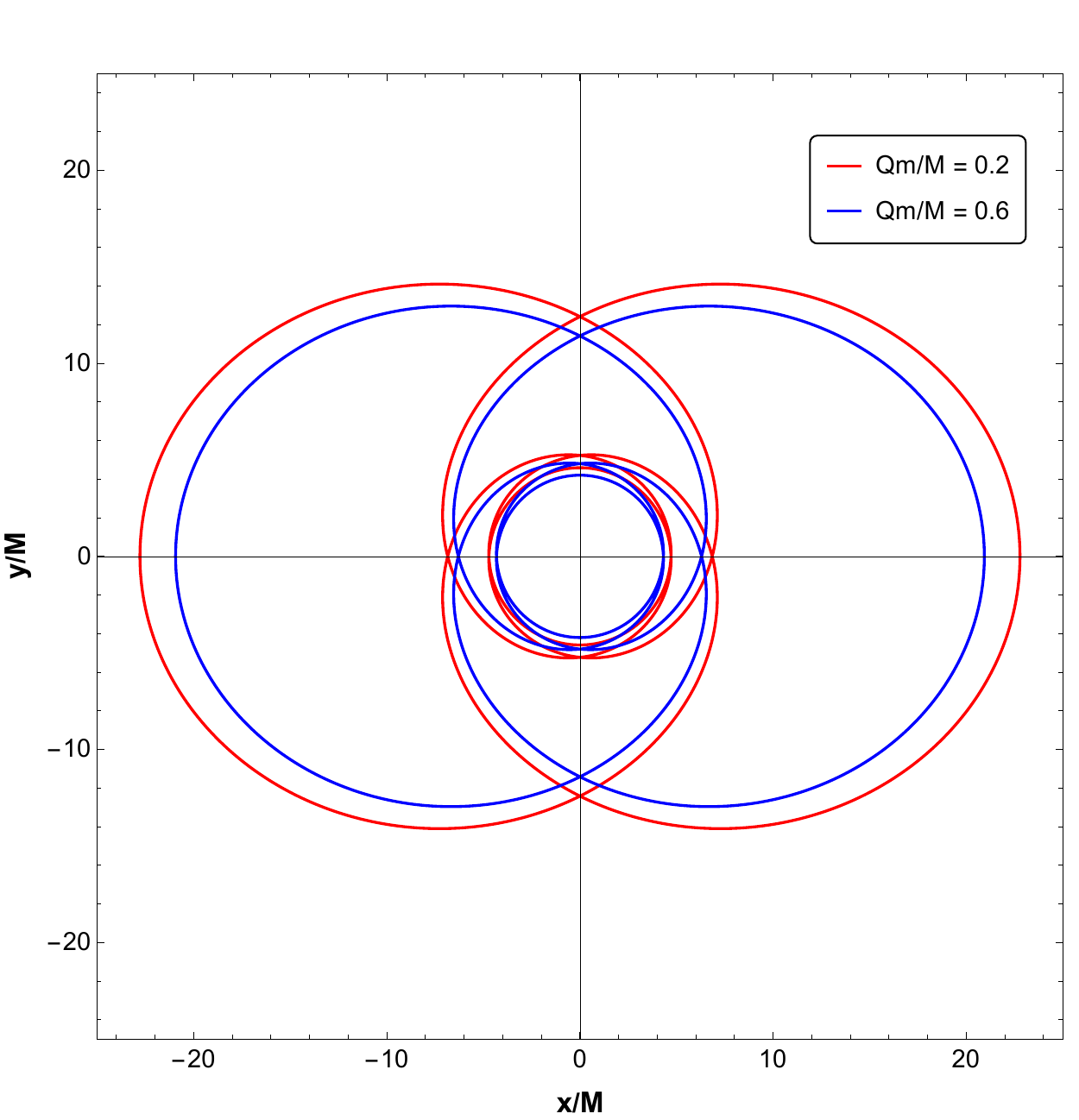}
			\put(50,85){\makebox(0,0){\small $(2, 1, 1)$}}
		\end{overpic}
	\end{minipage}
	\hfill
	\begin{minipage}{0.32\textwidth}
		\centering
		\begin{overpic}[width=\linewidth]{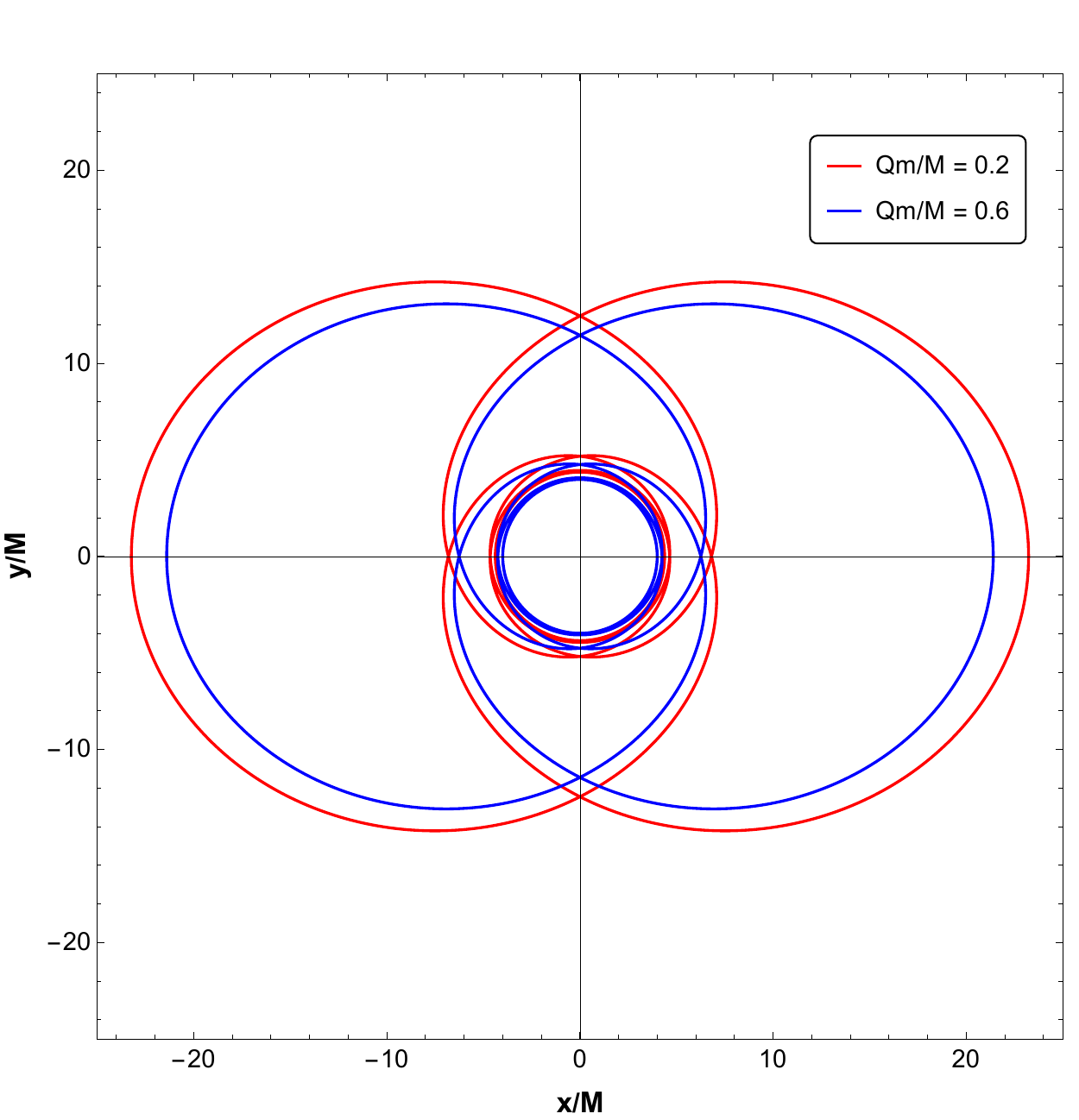}
			\put(50,85){\makebox(0,0){\small $(2, 2, 1)$}}
		\end{overpic}
	\end{minipage}
	\hfill
	\begin{minipage}{0.32\textwidth}
		\centering
		\begin{overpic}[width=\linewidth]{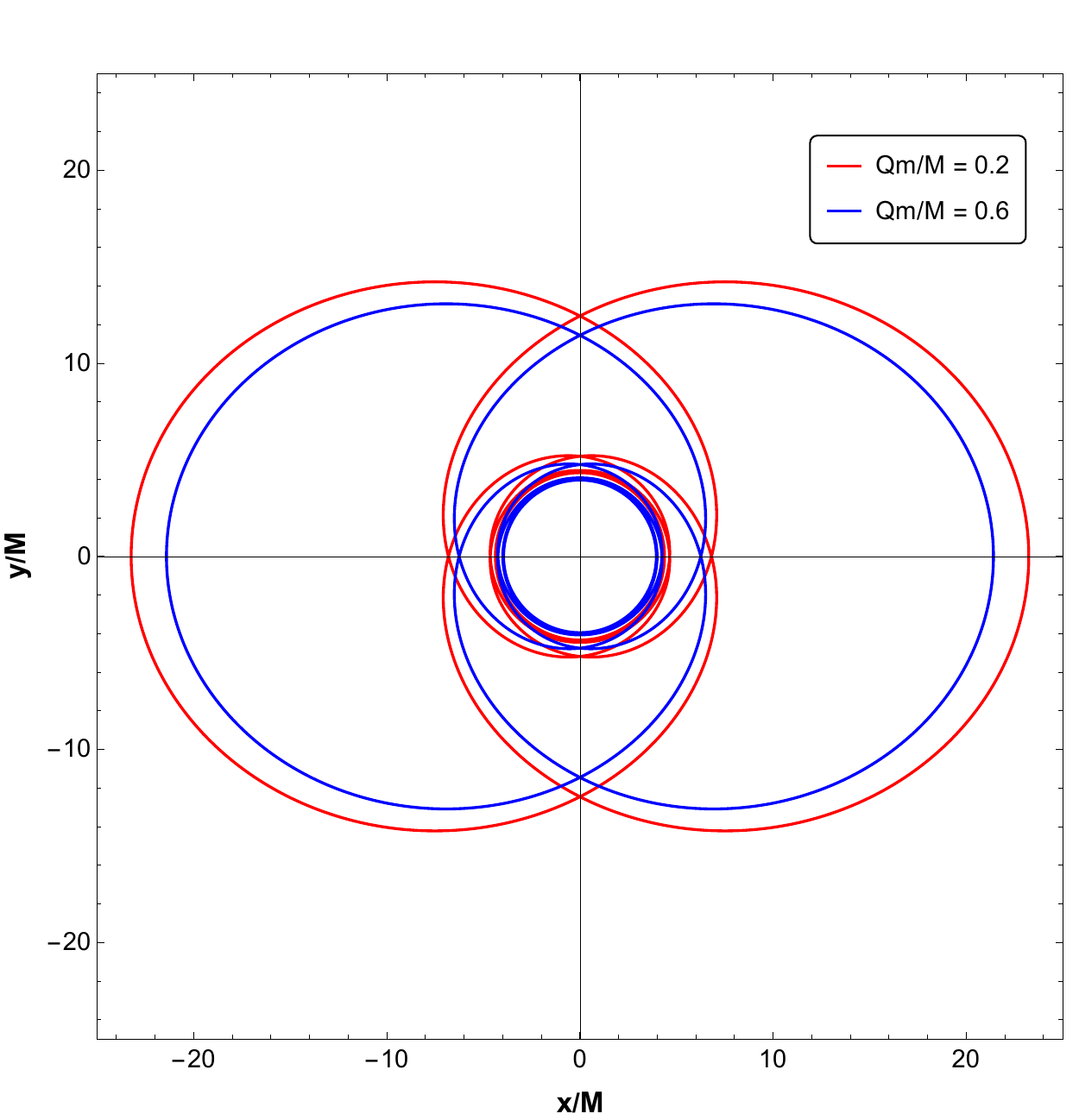}
			\put(50,85){\makebox(0,0){\small $(2, 3, 1)$}}
		\end{overpic}
	\end{minipage}
	\hfill
	\vspace{0.2cm}
	
	\begin{minipage}{0.32\textwidth}
		\centering
		\begin{overpic}[width=\linewidth]{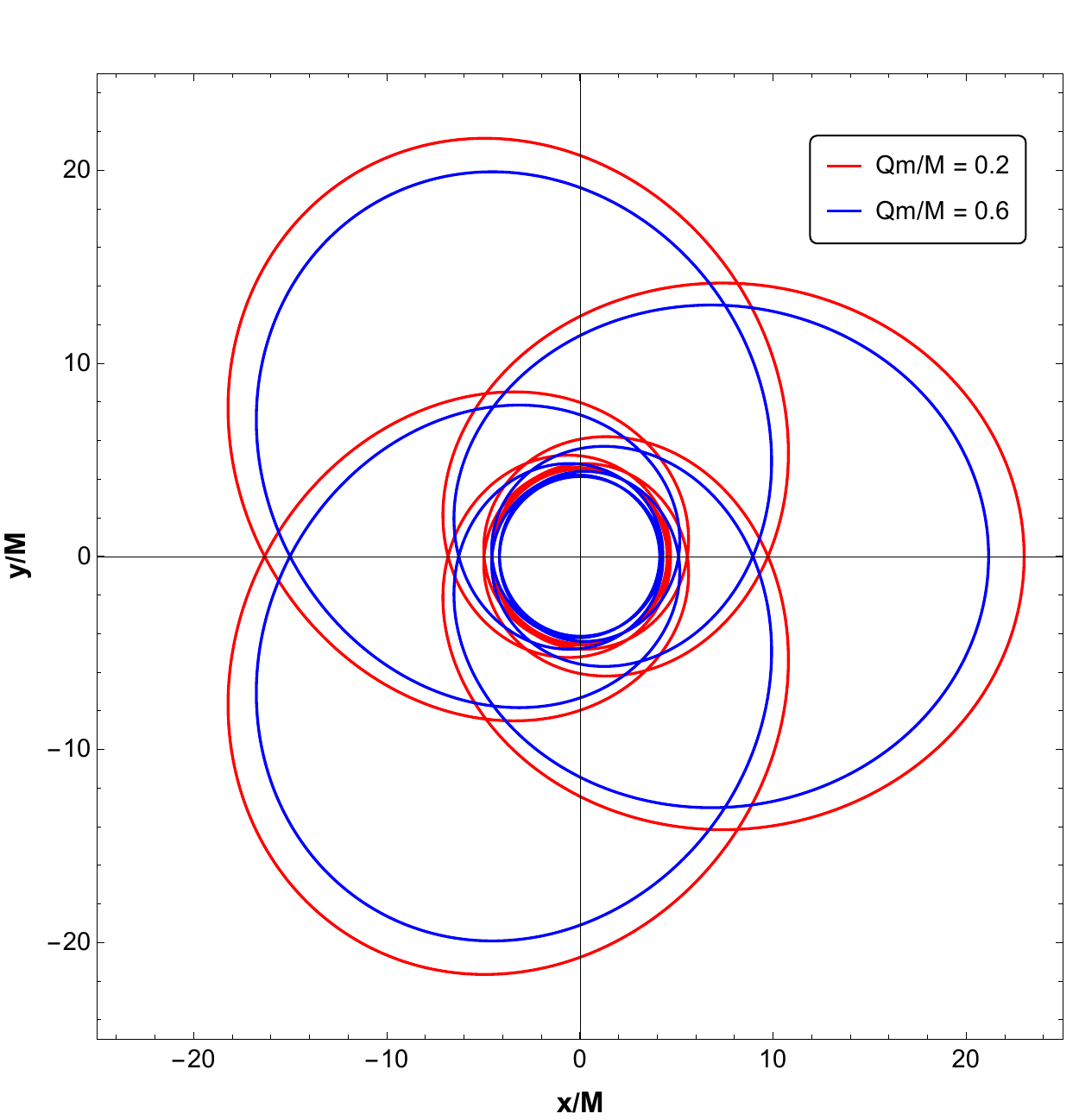}
			\put(50,85){\makebox(0,0){\small $(3, 1, 2)$}}
		\end{overpic}
	\end{minipage}
	\hfill
	\begin{minipage}{0.32\textwidth}
		\centering
		\begin{overpic}[width=\linewidth]{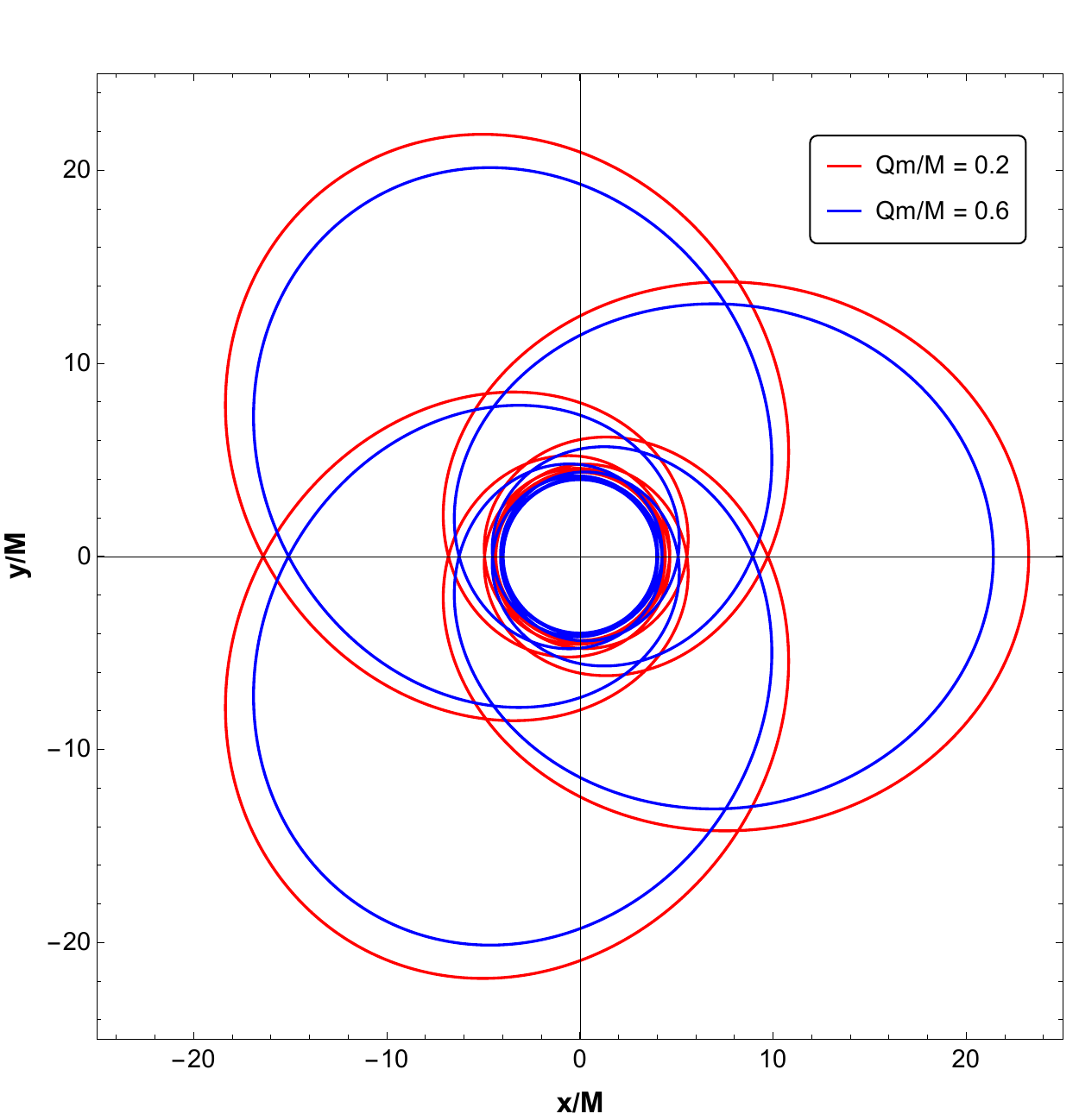}
			\put(50,85){\makebox(0,0){\small $(3, 2, 2)$}}
		\end{overpic}
	\end{minipage}
	\hfill
	\begin{minipage}{0.32\textwidth}
		\centering
		\begin{overpic}[width=\linewidth]{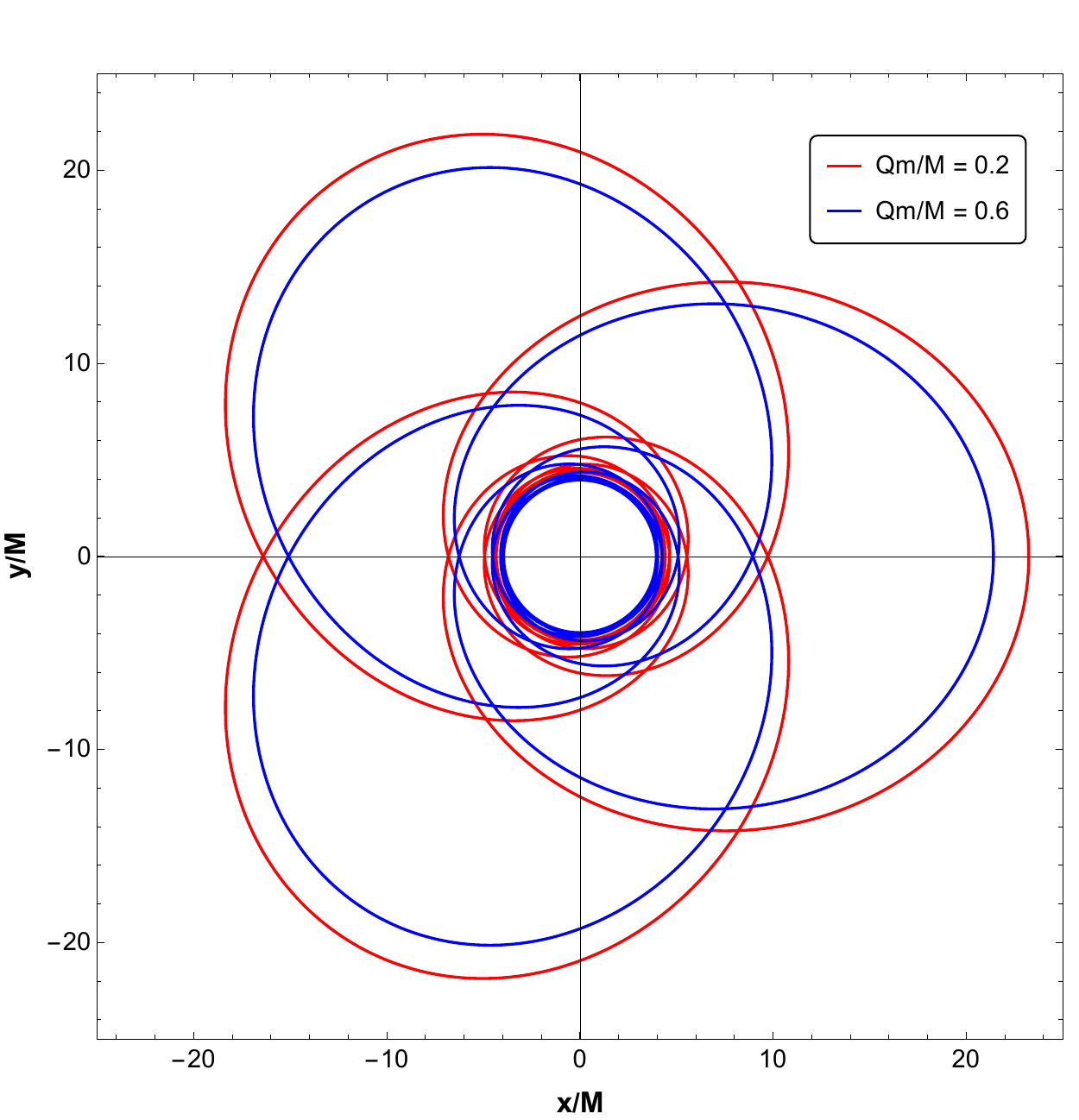}
			\put(50,85){\makebox(0,0){\small $(3, 3, 2)$}}
		\end{overpic}
	\end{minipage}
	\hfill
	\vspace{0.2cm}
	\caption{Periodic orbits characterized by different configurations $(z,w,v)$ for fixed angular momentum $L=\frac{1}{2}(L_{\rm ISCO}+L_{\rm MBO})$. The red image and the blue image respectively represent $Q_m/M=0.2$ and $Q_m/M=0.6$. The nonlinear electromagnetic parameter and coupling strength are fixed at $a/M^2=5$ and $\beta=0.4$.
	For each configuration, the energy values are selected from Table \ref{tab:energy_q} to satisfy $q=w+\frac{v}{z}$.}
	\label{fig:orbits_fixedL}
\end{figure*}

In addition to the periodic orbit analysis for the fixed orbital energy (e.g., $E=0.96$), we can also provide a similar discussion on the periodic orbits for a given angular momentum. As a representative example, we select the angular momentum at the arithmetic mean of the angular momenta associated with the ISCO and MBO
\begin{equation}\label{L_fixed}
    L=\frac{L_{\rm ISCO}+L_{\rm MBO}}{2}.
\end{equation}
This choice places the small compact object in a strong-field regime while preserving a potential well that supports stable bound motion. 
For illustration, we take $a/M^2=5$ and $\beta=0.4$, for which the numerical results yield $q_{\min}<1$ (so that periodic orbits with $w \ge 1$ can be obtained). 
Under the current parameter choices, the corresponding energy values for different orbital configurations $(z,w,v)$ and magnetic charges are summarized in Table~\ref{tab:energy_q}.

To display the influence of magnetic charge on periodic orbits, the orbit trajectories in fixed angular momentum cases for different magnetic charges are shown in Fig.~\ref{fig:orbits_fixedL}. For larger $Q_m$, a reduction of the apoapsis has been witnessed, and the area of the whirl becomes smaller. Similar behavior is also found for other model parameter values $(a,\beta)$ within the parameter range considered in this work. Notably, a comparison of Fig.~\ref{fig:orbits_fixedE} and Fig.~\ref{fig:orbits_fixedL} reveals that the magnetic charge affects the apastron radius in opposite ways for fixed energy (e.g., $E=0.96$) versus fixed angular momentum (e.g., $L=\frac{1}{2}(L_{\rm MBO} + L_{\rm ISCO})$). For fixed energy cases, the apastron radius increases with increasing $Q_m$, whereas for fixed angular momentum cases, it decreases as $Q_m$ increases.

The periodic trajectories obtained under the fixed-energy and fixed-angular-momentum conditions provide a useful avenue for studying the electromagnetic black hole systems in the strong-field region and their gravitational radiation signatures. In the next section, we further explore the gravitational radiation emitted by a small compact object moving along these trajectories and investigate how the electromagnetic field modifies the resulting waveforms.

%\clearpage

\section{gravitational waveforms}\label{sec4}

In this section, we investigate the gravitational wave signals radiated from these periodic orbits by a stellar-mass compact object orbiting the supermassive magnetically charged black hole in $f(R,T)$ gravity. We employ the ``Numerical Kludge'' method to analyse gravitational waveforms, whose main strategy consists of two key steps: first, calculate the orbit of the small object by solving the geodesic equation; second, use the quadrupole formula of gravitational radiation to get the corresponding gravitational waves. This approach is widely used to generate waveforms for various kinds of bound orbits in EMRIs \cite{Tu:2023xab,He:2026txc}.

We consider a small object with mass $m$ orbiting a central black hole with mass $M$ (where $m \ll M$). The gravitational waves for such systems can be treated as perturbations $h_{ij}$ on the background spacetime (generated by the central massive black hole). In the quadrupole approximation, the spatial components of the metric perturbation in the trace-reversed gauge are given by the second time derivative of the quadrupole moment tensor $I_{ij}$
\begin{eqnarray}\label{h_{ij}}
	h_{ij} = \frac{2}{D_L} \frac{d^2 I_{ij}}{dt^2},
\end{eqnarray}
where $D_L$ is the luminosity distance to the source. We can conveniently treat the small compact object in EMRIs as a point mass moving on a trajectory $x_i(t)$, so the quadrupole moment tensor can be defined as $I_{ij} = m x_i(t) x_j(t)$. Substituting this quadrupole moment tensor into Eq.~(\ref{h_{ij}}),
one obtains
\begin{equation}
h_{ij}(t)
=
\frac{4\xi M}{D_L}
\left[
v_i(t)v_j(t)
-
\frac{m}{r(t)} n_i(t)n_j(t)
\right],
\end{equation}
where $v_i(t)=\frac{dx_i(t)}{dt}$ is the velocity of the compact object,
$n_i(t)=\frac{x_i(t)}{r(t)}$ is the radial unit vector, and $\xi=\frac{mM}{(m+M)^2}$ is the symmetric mass ratio.

\begin{figure}
	\centering
	\includegraphics[width=0.8\textwidth]{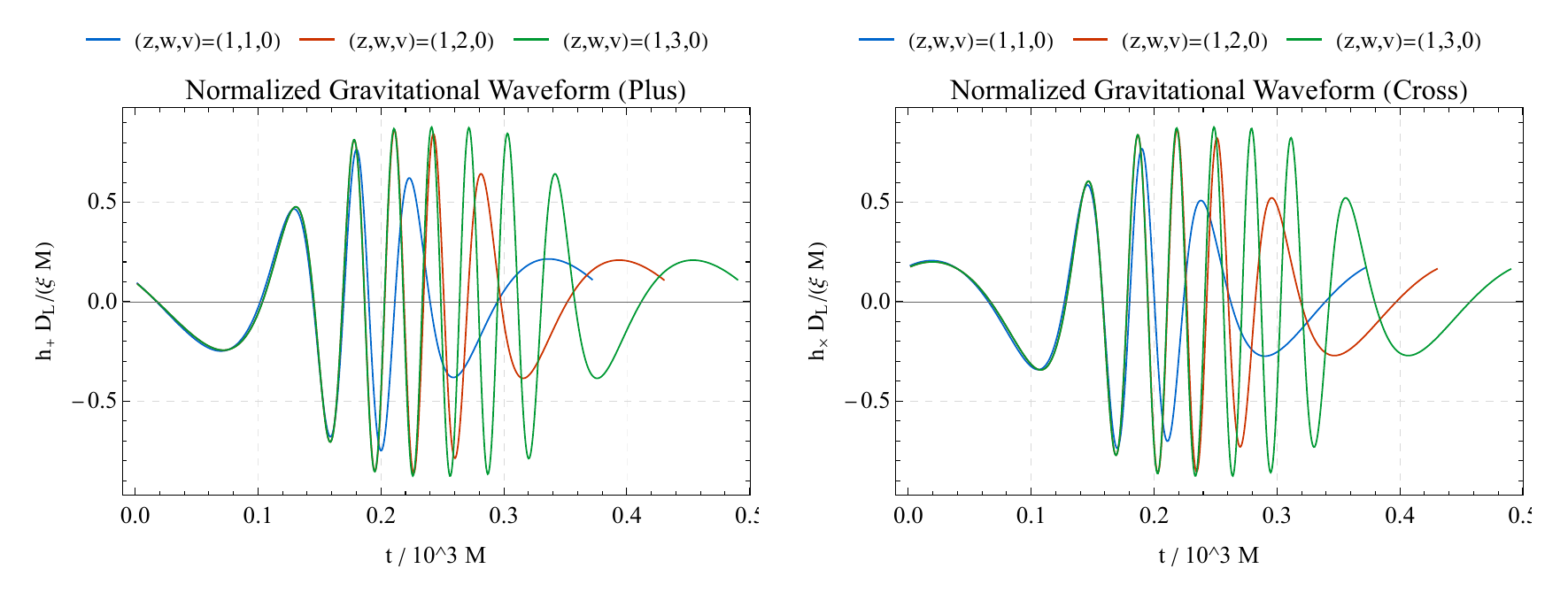}\\[0.8em]
	\includegraphics[width=0.8\textwidth]{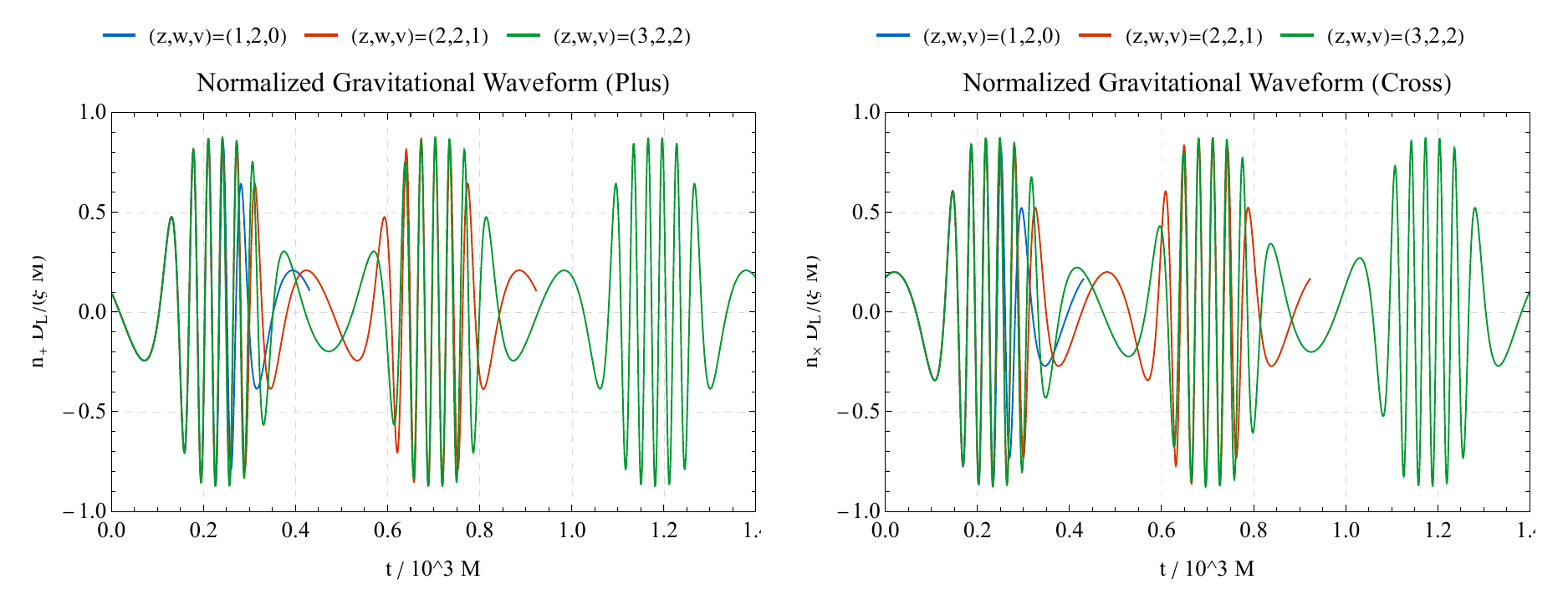}
	\caption{
		Gravitational waveforms generated by a compact object moving along periodic orbits around the magnetically charged black hole. 
		In this figure, the left and right plots show the plus polarization $h_{+}$ and the cross polarization $h_{\times}$, respectively. 
		For convenience, the amplitude on the vertical axis is normalized by a factor of $\xi M$. 
		The upper panels compare the gravitational waveforms for different configurations $(z,w,v)=(1,1,0), (1,2,0), (1,3,0)$. 
		The lower panels show the gravitational waveforms corresponding to different periodic-orbit configurations $(z,w,v)=(1,2,0), (2,2,1), (3,2,2)$, illustrating the diversity of waveform morphologies associated with different periodic orbits. 
		In this figure, the black-hole parameters are fixed at $Q_m/M=0.2$, $a/M^2=5.0$, $\beta=0.4$, while the orbital energy is fixed at $E=0.96$. 
		The small compact-object mass is chosen such that the mass ratio is $m/M=10^{-5}$ and the symmetric mass ratio is $\xi=mM/(m+M)^2\approx10^{-5}$. 
		The luminosity distance is fixed at $D_L=10^{15}M$, and the viewing azimuthal angle is taken as $\zeta=2\pi/3$. 
	}
	\label{fig:waveforms_q}
\end{figure}

To calculate the waveform observed at the detector, we first transform the particle's trajectory from the spherical coordinates $(r, \theta, \phi)$ to the Cartesian coordinates $(x, y, z)$ of the source frame. Then we can calculate the perturbation tensor $h_{ij}$ from the orbit trajectory $(x(t),y(t),z(t))$.
The gravitational wave signals are typically decomposed into two independent polarization modes: plus ($h_+$) and cross ($h_\times$). To obtain these, we construct a detector-adapted coordinate system $(X, Y, Z)$ centered on the black hole, where the $Z$-axis points towards the observer. The orientation of this frame is defined by the inclination angle $\iota$ (the angle between the orbital angular momentum and the line of sight) and the azimuthal angle $\zeta$. The basis vectors of the detector frame $(X,Y,Z)$ can be expressed in terms of the source frame $(x,y,z)$ as
\begin{eqnarray}
	\mathbf{e}_X &=& (\cos \zeta, -\sin \zeta, 0), \nonumber\\
	\mathbf{e}_Y &=& (\cos \iota \sin \zeta, \cos \iota \cos \zeta, -\sin \iota), \\
	\mathbf{e}_Z &=& (\sin \iota \sin \zeta, \sin \iota \cos \zeta, \ \ \ \cos \iota). \nonumber
\end{eqnarray}
Projecting the metric perturbation tensor onto this basis, the two polarization states are given by \cite{Babak:2006uv}
\begin{eqnarray}
	h_+ &=& \frac{1}{2} \big(\mathbf{e}_X^i \mathbf{e}_X^j - \mathbf{e}_Y^i \mathbf{e}_Y^j\big) h_{ij}, \nonumber\\
	h_\times &=& \frac{1}{2} \big(\mathbf{e}_X^i \mathbf{e}_Y^j + \mathbf{e}_Y^i \mathbf{e}_X^j\big) h_{ij}.
\end{eqnarray}
For the numerical results presented below, we assume the observer is located at a distance $D_L$ with an inclination angle $\iota = 0$ and azimuthal angle $\zeta = \frac{2\pi}{3}$.

We first analyze how the topological structure of the periodic orbits, which is inherent in their configurations $(z,w,v)$, affects the gravitational waveforms. Fig.~\ref{fig:waveforms_q} displays the gravitational waveform $h_+$ and $h_{\times}$ for periodic orbits characterized by different orbital configurations $(z,w,v)$ at the given orbital energy $E=0.96$. The angular momentum L is determined separately via $q=w+\frac{v}{z}$ to satisfy the corresponding periodic-orbit structure. 
For simple configurations (e.g., $(z,w,v)=(1,1,0)$), the waveform exhibits a relatively simple structure. 
However, the richness of the waveform structure increases with the complexity of the orbit $(z, w, v)$. The ``zoom-whirl'' characteristics of periodic orbits are imprinted on their gravitational waveform as a complex modulation, where high-frequency bursts (corresponding to the periastron passage where velocity is highest) are separated by lower-frequency intervals (corresponding to the apastron). Specifically, as the zoom number $z$ increases (e.g., $(z,w,v)=(1,2,0)$, $(2,2,1)$, and $(3,2,2)$), the gravitational waveform exhibits a larger number of high-frequency burst stages and low-frequency stages. As the whirl number $w$ increases (e.g., $(z,w,v)=(1,1,0)$, $(1,2,0)$ and $(1,3,0)$), the duration of the high-frequency burst stages becomes longer, accompanied by an increase in the number of oscillations during these stages. This behavior is inherited by the orbital structure of periodic trajectories; for example, for the configurations $w>1$, the small compact object undergoes repeated whirl motion around the black hole within each radial period, giving rise to the longer-term high-frequency features in the waveform.
The above results indicate the richness of periodic orbits' waveforms, providing a unique fingerprint for their orbital structure.

\begin{figure}
    \centering
    \includegraphics[width=0.8\textwidth]{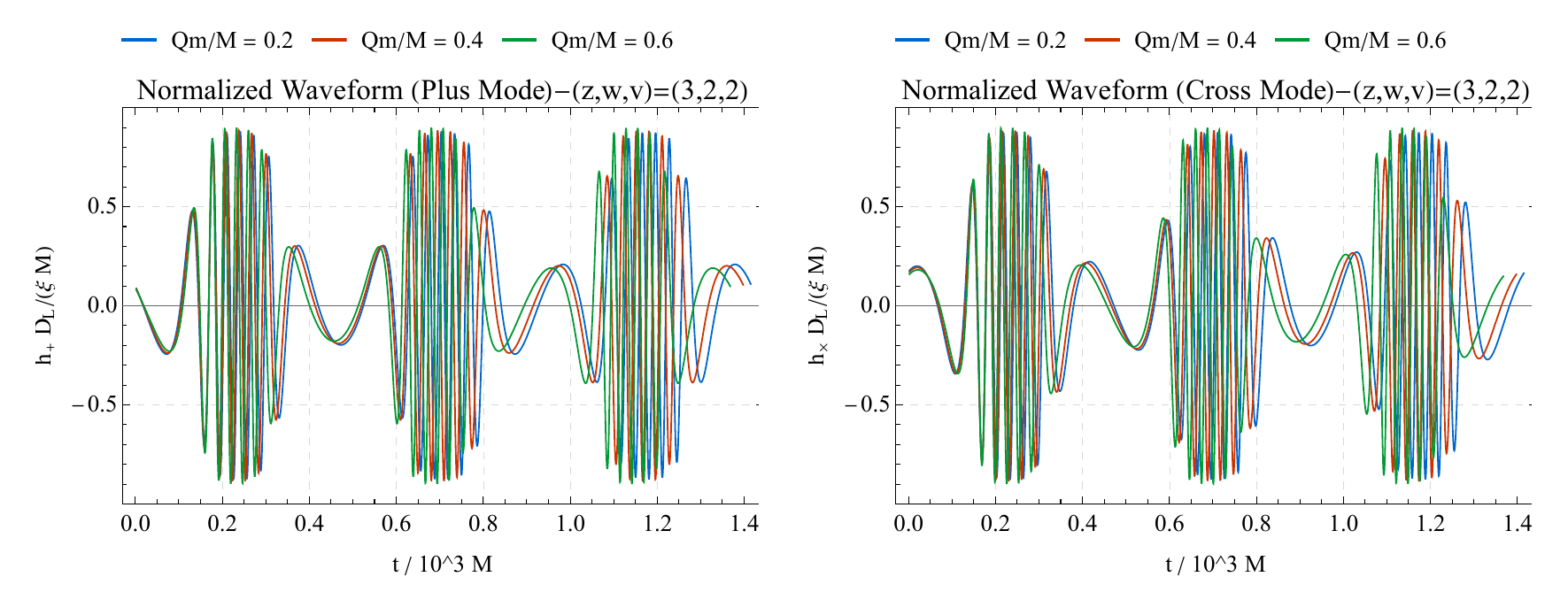}
    \caption{
    Comparison of gravitational waveforms generated by a compact object moving along the specific periodic orbit $(z,w,v)=(3,2,2)$ under different magnetic charge $Q_m$. 
    The left and right panels show the plus polarization $h_{+}$ and the cross polarization $h_{\times}$, respectively. 
    For convenience, the amplitude on the vertical axis is normalized by a factor of $\xi M$. 
    The three curves correspond to $Q_m/M=0.2$, $Q_m/M=0.4$, and $Q_m/M=0.6$. 
    In this figure, the Euler-Heisenberg parameter and the $f(R,T)$ coupling parameter are fixed at $a/M^2=5.0$ and $\beta=0.4$, respectively. 
    We select the small compact-object mass such that the mass ratio is $m/M=10^{-5}$ and the symmetric mass ratio is $\xi=mM/(m+M)^2\approx10^{-5}$. 
    The luminosity distance is fixed at $D_L=10^{15}M$, and the viewing azimuthal angle is taken as $\zeta=2\pi/3$. 
    }
    \label{fig:waveform322}
\end{figure}

\begin{figure}
	\centering
	\includegraphics[width=0.95\textwidth]{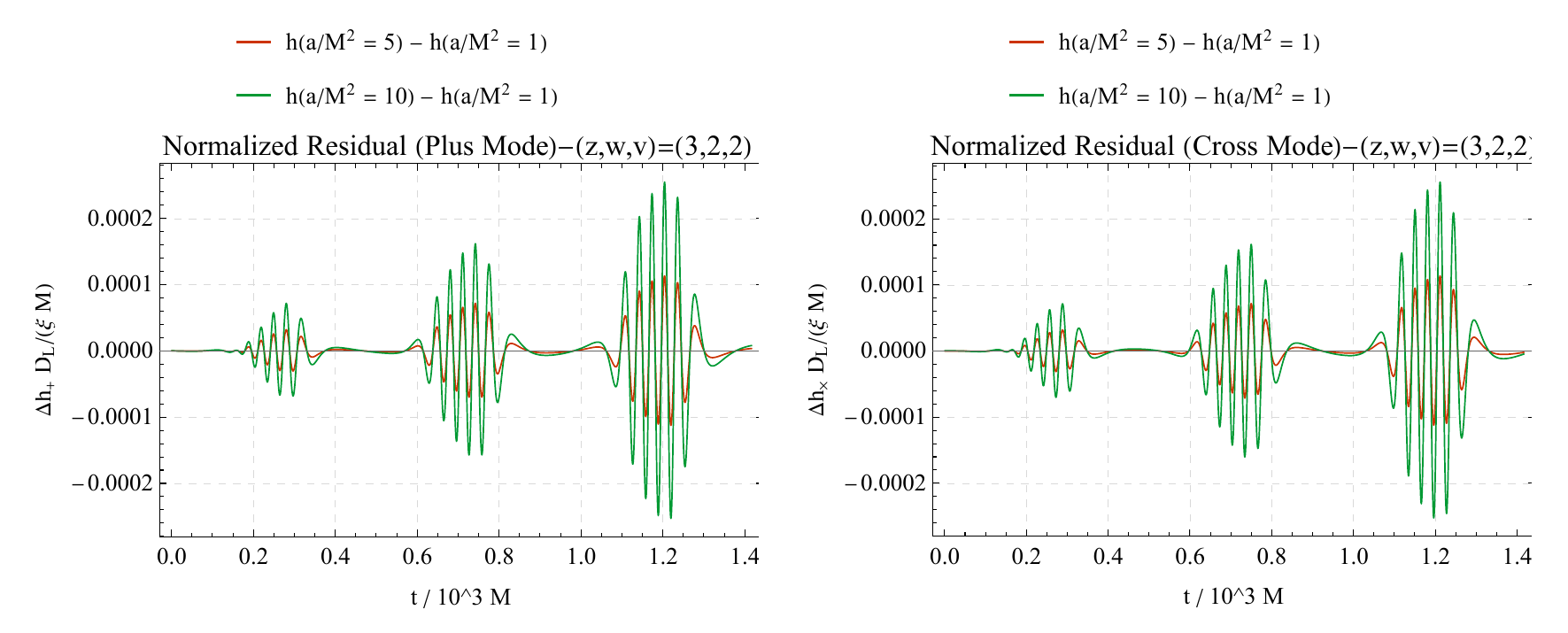}
	\caption{Normalized residual of gravitational waveforms for different nonlinear electrodynamics parameters $a$ with fixed $E=0.96$, $Q_m/M=0.2$, $\beta=0.4$, and periodic-orbit configuration $(z,w,v)=(3,2,2)$, corresponding to $q=8/3$. For each value of $a$, the angular momentum $L$ is determined separately to satisfy the same periodic-orbit structure. The residuals are calculated with respect to the reference waveform for $a_{\text{ref}}/M^2=1$, namely, $\Delta h=h(a)-h(a_{\text{ref}})$. The red and green curves correspond to $\Delta h=h(a)-h(a_{\text{ref}})$ for $a/M^2=5$ and $a/M^2=10$, respectively. The left and right panels show the normalized residuals of the plus and cross polarizations, respectively.}
	\label{fig:waveform residual}
\end{figure}

Next, we investigate the influence of the magnetic charge $Q_m$ on the gravitational wave. In Fig.~\ref{fig:waveform322}, we plot the waveforms for a given periodic orbit configuration $(z, w, v) = (3, 2, 2)$ while varying the magnetic charge $Q_m$.
It is observed that the magnetic charge induces a phase shift and slightly modifies the amplitude of the gravitational wave signal. Although the orbit configuration $(z, w, v)$ is fixed, the apoapsis radius $r_a$ and the time period $T$ of the orbit change with $Q_m$ due to the modification of the background metric function $f(r)$. Then, the orbital frequencies of the motion are altered, leading to a noticeable dephasing in the waveform over time. 
This sensitivity of the waveform and its small variations induced by magnetic charge suggest that gravitational wave observations could potentially constrain the magnetic charge of black holes.

Besides the magnetic charge, we also investigate the influence of the nonlinear electrodynamics parameter $a$ on the gravitational waveforms. 
Since the effect of $a$ on the waveforms is relatively weak, we provide the relative waveform residuals $\delta h=h(a)-h(a_{\text{ref}})$, which is the variation of waveform relative to the reference case $a_{\text{ref}}/M^2=1$. The numerical results for waveform residuals under different $a$ are given in Fig.~\ref{fig:waveform residual}.
From the figure, we can see that the residuals caused by varying the nonlinear electrodynamics parameter $a$ are mainly concentrated in the high-frequency burst stages. 
The residual amplitude increases as $a$ increases, indicating a growing difference in the waveform. Moreover, the residual peaks become progressively larger in the later high-frequency stages, with the largest deviation appearing in the final burst.

\section{Conclusion}\label{sec5}

In this paper, we have presented a comprehensive study of the periodic motions of a small compact object orbiting a magnetically charged black hole and their gravitational radiation within the framework of $f(R,T)$ gravity coupled with Euler-Heisenberg electrodynamics. 
This study helps us understand how nonlinear electrodynamics and modified gravity influence the specific bound orbits in EMRI systems. 
The investigation of periodic-orbit structures and gravitational waveforms presented in this study provides theoretical guidance for further exploring and probing the strong gravitational field effects and electromagnetic effects in black hole systems.

Based on an analytical magnetically charged black hole solution, we analyze the effective potential of the test particles, from which several critical stable circular orbits are calculated and studied, including the innermost stable circular orbit (ISCO) and marginally bound orbit (MBO). From the numerical results of ISCO and MBO, we observed that the radii of these characteristic orbits are more sensitive to the magnetic charge $Q_m$ than the Euler-Heisenberg nonlinear parameter $a$.

The periodic orbits characterized by zoom-whirl-vertex integers $(z,w,v)$ are mainly focused on in our studies. We calculate the trajectories of periodic orbits and the corresponding precession parameter $q$. By exploring how orbital precession varies with energy $E$ and angular momentum $L$, our analysis reveals that the precession parameter $q$ possesses a minimum value for bound orbits, i.e., $q_{\text{min}} \le q < +\infty$. 
We also explain the physical origin of $q_{\text{min}}$ and provide a rigorous analytical derivation of this lower bound in terms of the azimuthal and radial epicyclic frequencies. 
Comparing the periodic orbits obtained using different black hole parameters, the influence of the magnetic charge $Q_m$ on the precession parameter and orbital trajectories is successfully extracted. 
For a fixed orbital energy or angular momentum, the precession parameter $q$ decreases as $Q_m$ increases, indicating that the azimuthal angle accumulated during one radial period is reduced. Additionally, the orbital trajectories are also significantly modified by the magnetic charge. 
For the same periodic-orbit configuration $(z,w,v)$, the apoapsis radius increases with $Q_m$ under the fixed-energy condition, whereas it decreases with $Q_m$ under the fixed-angular-momentum condition. The size of the whirl region is also modified accordingly. As the magnetic charge $Q_m$ increases, the area of the whirl region decreases.
These results demonstrate that the magnetic charge can significantly modify not only the precession parameter but also the orbital structure of the periodic orbits. However, the nonlinear electrodynamics parameter $a$ has a relatively weaker influence on the periodic-orbit trajectories, compared with the magnetic charge $Q_m$.

Finally, utilizing the numerical kludge method, we computed the gravitational waveforms emitted by these periodic orbits in EMRIs. 
Our results indicate that the waveform morphology is directly linked to the orbital taxonomy $(z, w, v)$, exhibiting distinct ``zoom-whirl'' features (composed of high-frequency bursts separated by low-frequency signals) during the zoom and whirl stages. 
Moreover, the comparison of waveforms for different magnetic charges reveals that $Q_m$ introduces a significant phase shift, providing a potential observational signature of the electrodynamic effects in black hole systems. 
Relative to the magnetic charge, other parameters have a weaker influence on the gravitational waves.
In summary, this work demonstrates that periodic orbits and their gravitational wave signatures can serve as a sensitive probe of electromagnetic and strong-field effects in black hole spacetimes, offering a valuable theoretical basis for future EMRI-based tests of modified gravity and nonlinear electrodynamics.

\section*{Acknowledgement}

The authors thank Yi-Zhi Liang, Mian Zhu, and Guo-He Li for helpful discussions. This research was funded by the Natural Science Foundation of China (Grant No. 12175212), Natural Science Foundation of Chongqing Municipality (Grant No. CSTB2022NSCQ-MSX0932), and the Scientific and Technological Research Program of Chongqing Municipal Education Commission (Grant No. KJQN202201126).

\appendix

% Replace the following bibliography with the bibliography file used in the
% complete manuscript. Undefined citations only produce warnings during a
% preliminary compilation of this standalone section.

\end{document}